\documentclass[prb,superscriptaddress,amsfonts,twocolumn,showpacs,floatfix]{revtex4}
\usepackage{amsmath}
\usepackage{amsfonts}
\usepackage{amssymb}
\usepackage{bm}
\usepackage{tabularx}
\usepackage{multirow}
\usepackage[dvips]{graphicx,color}

\newcommand{\TN}{T_{\mathrm{N}}}

\begin{document}
\title{
Magnetic structure of the conductive triangular-lattice antiferromagnet PdCrO$_{2}$
}

\author{Hiroshi Takatsu}
\affiliation{Department of Physics, Tokyo Metropolitan University, Hachioji-shi, Tokyo 192-0397, Japan}

\author{Gwilherm N$\acute{\mathrm{e}}$nert}
\affiliation{Institut Laue-Langevin, 6 rue Jules Horowitz, BP 156,38042 Grenoble Cedex 9, France}

\author{Hiroaki Kadowaki}
\affiliation{Department of Physics, Tokyo Metropolitan University, Hachioji-shi, Tokyo 192-0397, Japan}

\author{Hideki Yoshizawa}
\affiliation{Neutron Science Laboratory, Institute for Solid State Physics, The University of Tokyo, Tokai Ibaraki 319-1106, Japan}

\author{Mechthild Enderle}
\affiliation{Institut Laue-Langevin, 6 rue Jules Horowitz, BP 156,38042 Grenoble Cedex 9, France}

\author{Shingo Yonezawa}
\affiliation{Department of Physics, Graduate School of Science, Kyoto University, Kyoto 606-8502, Japan}

\author{Yoshiteru Maeno}
\affiliation{Department of Physics, Graduate School of Science, Kyoto University, Kyoto 606-8502, Japan}

\author{Jungeun Kim}
\affiliation{Japan Synchrotron Radiation Research Institute/SPring-8, 1-1-1 Kouto, Sayo, Hyogo 679-5198, Japan}

\author{Naruki Tsuji}
\affiliation{Japan Synchrotron Radiation Research Institute/SPring-8, 1-1-1 Kouto, Sayo, Hyogo 679-5198, Japan}

\author{Masaki Takata}
\affiliation{Japan Synchrotron Radiation Research Institute/SPring-8, 1-1-1 Kouto, Sayo, Hyogo 679-5198, Japan}

\author{Yang Zhao}
\affiliation{Department of Physics and Astronomy, Johns Hopkins University, Baltimore, Maryland 21218, USA}
\affiliation{NIST Center for Neutron Research, National Institute of Standards and Technology, Gaithersburg, Maryland 20899, USA }
\affiliation{Department of Materials Science and Engineering, University of Maryland, College Park, Maryland 20742, USA}

\author{Mark Green}
\affiliation{NCNR, National Institute of Standards and Technology, Gaithersburg, MD 20899-6102, U.S.A.}

\author{Collin Broholm}
\affiliation{Department of Physics and Astronomy, Johns Hopkins University, Baltimore, Maryland 21218, USA}
\date{\today}

\begin{abstract}
We performed neutron single crystal and synchrotron X-ray powder diffraction
experiments  
in order to investigate the magnetic and crystal structures of the conductive 
layered triangular-lattice antiferromagnet PdCrO$_2$ with a putative spin chirality,
which contributes to an unconventional anomalous Hall effect. 
We revealed that the ground-state magnetic structure is a commensurate and nearly-coplanar 120$^\circ$ spin structure.
The 120$^\circ$ plane in different Cr layers seem to tilt with one another, leading to a small non-coplanarity.
Such a small but finite non-coplanar stacking of the 120$^\circ$ planes 
gives rise to a finite scalar spin chirality,
which may be responsible for the unconventional nature of the Hall effect of PdCrO$_2$.
\end{abstract}


\maketitle

\section{Introduction}
Recently, there has been a rapid progress in the study of novel magneto-electric phenomena, 
such as magnetic multiferroics and unconventional anomalous Hall 
effect (UAHE)~\cite{N.Nagaosa2010,D.Xiao2010,TokuraAM2010,ArimaJPSJ2011}. 
Common to all of these is that they involve a spin current, 
i.e., magnetic structures with spin chiralities.
In case of UAHE, a topological quantum effect has been proposed as a potential mechanism~\cite{N.Nagaosa2010,D.Xiao2010}:
In a magnetic structure with the scalar spin chirality $\chi_{ijk} = \bm{S}_i\cdot(\bm{S}_j \times \bm{S}_k)$,
the wave function of a conduction electron gains a Berry phase, which 
plays a role of a fictitious magnetic field and leads to appearance of the Hall voltage even
without the net magnetization.
The magnitude of the fictitious field is proportional to the solid angle formed by 
the three non-coplanar spins~\cite{Taguchi2001,Y.YasuiJPSJ2006,Machida2007,Y.MachidaNature2009,P.Matl1998PRB, J.YePRL1999,Ohgushi2000,TataraJPSJ2002,T.TomizawaPRB2009,K.Taguchi2009}.
This mechanism is in analogy to the Aharonov-Bohm effect~\cite{Aharonov1959}.

In search of UAHE attributable to the spin chirality,
geometrically frustrated magnets are particularly promising, 
because they often exhibit non-coplanar spin configurations with finite spin chiralities. 
Indeed, the UAHE has been observed in materials with structures that are three-dimensional analogues of 
the triangular lattice (TL)~\cite{Taguchi2001,Y.YasuiJPSJ2006,Machida2007,Y.MachidaNature2009,P.Matl1998PRB}.
However, in two dimensional (2D) TL systems, 
which is the simplest example of a geometrically frustrated spin system, 
the UAHE has been observed only recently~\cite{H.TakatsuPRL2010,Y.Shiomi2012}.
Naively, the UAHE driven by the Berry-phase concept cannot be expected in a coplanar 
120$^\circ$ spin structure, which is often realized in 2D-TL antiferromagnets.
This is because $\chi_{ijk}$ is locally zero for every triangles or,
even if $\chi_{ijk}$ is locally finite, the net chirality vanishes because 
$\chi_{ijk}$ of different triangles cancels out~\cite{N.Nagaosa2010,TataraJPSJ2002}.
However, $\chi_{ijk}$ may remain finite, 
if the spin chirality and magnetization are coupled with the help of the spin-orbit interaction~\cite{TataraJPSJ2002,Kawamura2003},
or in {\color[rgb]{0,0,0}non-coplanar spin structures with a four-site magnetic unit cell~\cite{ShindouPRL2001,I.Martin2008,Y.Akagi2010},
both of which change the balance of the uniform $\chi_{ijk}$ on each triangle.}
The fundamental mechanism for the UAHE observed in the frustrated 2D-TL systems is thus not well 
understood and is awaited to be clarified.

The delafossite compound PdCrO$_2$ is a rare example of a 2D-TL antiferromagnet that exhibits UAHE~\cite{H.TakatsuPRL2010}.
The metallic conduction of this material is predominantly attributed to the Pd $4d$ electron band~\cite{OkPRL2013,SobotaPRB2013,K.P.Ong2011},
and the magnetic properties are governed by the localized spins of Cr$^{3+}$ ions ($S=3/2$),
which order antiferromagnetically at $\TN=37.5$~K~\cite{Doumerc1986,Mekata1995,H.TakatsuPRB2009}.
The spin Hamiltonian of this system is approximately written as 
\begin{equation}
H = - 2J \sum_{<i,j>}\bm{S}_{i}\cdot\bm{S}_{j} - 2J'\sum_{<l,m>}\bm{S}_{l}\cdot\bm{S}_{m} + D\sum_{i}(S_{i}^z)^2,
\label{eq.0}
\end{equation}
where $J (<0)$ and $J'$ are the nearest-neighbor intraplane and interplane interactions, respectively,
and $D$ is the single ion anisotropy.
The anisotropy of the magnetic susceptibility $\chi$, associated with a sharp drop in $\chi_c$ below $\TN$,
strongly suggests an easy-axis anisotropy along the $z$ axis, $D<0$~\cite{K.P.Ong2011,T.HironeJPSJ1957}.
Remarkably, 
this material exhibits UAHE below $T^{*}\simeq20$~K, noticeably lower than $\TN$~\cite{H.TakatsuPRL2010}:
The Hall resistivity $\rho_{xy}$ exhibits an unusual non-linear field dependence.
Apparently it deviates from the conventional behavior that is a linear function of
both magnetic induction and magnetization~\cite{C.M.Hurd},
since the magnetization of PdCrO$_2$ is proportional to $H$ down to 2~K~\cite{H.TakatsuPRL2010}.
We expect that a non-coplanar spin structure with a finite spin chirality
probably plays a crucial role for the emergence of the UAHE in this compound.

In this study, we have performed neutron scattering experiments on a single crystalline sample of PdCrO$_2$
to determine the magnetic structure in zero magnetic field.
We found that it is a commensurate 120$^\circ$ spin structure,
and that a small change of the magnetic structure occurs around $T^*$.
The magnetic structure analysis suggests alternative stacking of 120$^\circ$ spin layers,
which seems to be tilt with one another. 
We thus identify a non-coplanar 120$^\circ$ spin structure as the probable origin of the UAHE,
because the scalar spin chirality mechanism 
will work in this structure in the presence of a net magnetization induced by an external magnetic field. 

\section{Experimental}
Single crystals of PdCrO$_2$ were grown by a NaCl flux using PdCrO$_2$ powder synthesized via a solid state reaction~\cite{H.Takatsu2010}.
Synchrotron X-ray powder diffraction experiments were performed on the BL02B2 beam line at SPring-8
from 300~K to 11~K. We used a powder sample prepared by crushing single crystals.
The powder was packed into a glass tube ($\phi= 0.1$~mm) and 
mounted into a closed-cycle $^4$He-gas refrigerator.
The wavelength of the incident beam was $\lambda=0.6$~\AA.
A homogeneous granularity of the sample was checked
by a homogeneous intensity distribution in the Debye-Scherrer diffraction rings.

Neutron single-crystal diffraction experiments were performed with 
the triple-axis spectrometers 4G and C11 installed at the research reactor JRR-3M at Japan Atomic Energy Agency.
The neutron wavelength was fixed at either $\lambda = 1.64$ or $2.35$~\AA\, (4G),
and at $4.07$~\AA\, (C11).
Pyrolytic graphite (PG) (002) reflections were used as both monochromator and analyzer.
Higher-order neutrons were removed by a PG-filter or a Be-filter.
We employed collimations 20'-20'-20'-20' (4G) or 20'-20'-open (C11).
The sample was mounted in a closed-cycle $^4$He-gas refrigerator
so that the horizontal scattering plane of the spectrometer coincided with the 
hexagonal ($h$ $h$ $l$) or ($h$ $k$ 0) zones of the $R\bar{3}m$ symmetry.
A precise determination of the crystal structure symmetry is described later.
Integrated intensities of many Bragg reflections were 
measured with the four circle diffractometer D10 at Institute Laue-Langevin (ILL). 
Incident neutrons of wavelength $\lambda = 2.36$~\AA\, monochromated by PG(002) were used.
The sample was mounted in a He flow cryostat.
In order to perform a detailed structural analysis,
experiments were also carried out at room temperature (RT) with 
the hot-neutron four-circle diffractometer D9 at ILL.
We used a neutron wavelength of $\lambda = 0.838$~\AA.
In all the experiments, we used the same single crystal that has 
the dimensions $1.5\times3.0\times0.2$~mm$^3$ 
with the flat plane shape along the hexagonal $ab$ plane.

\section{Results}
\subsection{Determination of the crystal structure}
\begin{figure}[t]
\begin{center}
 \includegraphics[width=0.44\textwidth]{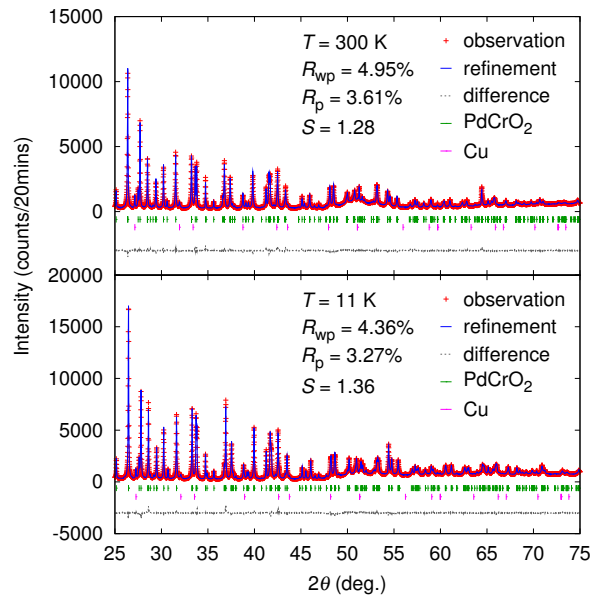}
\caption{
(Color online)
Synchrotron X-ray diffraction patterns of PdCrO$_2$ measured at 
$T=300$~K and 11~K. Observed and refined data are shown by 
crosses and solid curves, respectively.
The difference between the data and the model is plotted by the dashed curves in the lower part. 
Vertical bars represent positions of the Bragg reflections.
Additional Cu peaks come from the sample holder used in the experiments.
The Rietveld refinement revealed that PdCrO$_2$
retains to have the $R\bar{3}m$ crystal-structure symmetry down to 11~K. 
}
\label{fig.7}
\end{center}
\end{figure}
Since a precise crystal structure determination of PdCrO$_2$ has not been reported,
we have undertaken single crystal neutron and powder X-ray characterization.
%
Figure~\ref{fig.7} shows powder X-ray diffraction patterns taken at 300~K and 11~K.
The diffraction patterns were reasonably fitted by parameters of the delafossite structure with the space group 
$R\bar{3}m$ for both temperatures.
The $R$ factors of the Rietveld refinement~\cite{Izumi2007} were obtained as
$R_{\mathrm{wp}}=4.95$\%, $R_{\mathrm{e}}=3.87$\%, 
$R_{\mathrm{p}} = 3.61$\%, $R_{\mathrm{B}} = 4.94$\% for 300~K,
and 
$R_{\mathrm{wp}}=4.36$\%, $R_{\mathrm{e}}=3.20$\%, 
$R_{\mathrm{p}} = 3.27$\%, $R_{\mathrm{B}} = 4.70$\% for 11~K, respectively.
The goodness-of-fit parameter, $S = R_{\mathrm{wp}}/R_{\mathrm{e}}$,
was $S=1.28$ and 1.36 for 300~K and 11~K, respectively.
Excellent refinement was also confirmed by using neutron data at D9,
which achieved $\chi^{2}$ = 1.015 for 86 unique reflections.
The resulting structure parameters are listed in Tables~\ref{table_1} and ~\ref{table_1.2}.
These results demonstrate that PdCrO$_2$ remains in
the $R\bar{3}m$ symmetry down to low temperatures.

\begin{table}[t]
\begin{center}
 \caption{Structure parameters for PdCrO$_2$ refined by Rietveld analysis of  
 the X-ray data and the neutron data from D9.
 The analysis was performed assuming the space group $R\bar{3}m$ with atomic positions: 
 Pd 3a (0,0,0), Cr 3b (0,0,0.5), and O 6c (0,0,$z$). 
 $U_{\mathrm{iso}}$ represents the isotropic atomic displacement parameter.
 }
 \begin{tabular*}{0.46\textwidth}{@{\extracolsep{\fill}}lccc}
  \hline\hline
                  & x-ray(300 K)   & neutron(RT) & x-ray(11 K)            \rule{0mm}{4mm}   \\ \hline
   \multicolumn{4}{l}{{Cell parameters and positions}}                    \rule{0mm}{4.5mm} \\ 
   \,$a$ (\AA)    & \,\,2.9228(2)  &2.9280(1)    & \,\,2.9011(3)          \rule{0mm}{3.5mm} \\
   \,$c$ (\AA)    & 18.093(1)      &18.1217(9)   & 18.028(2)              \rule{0mm}{3.5mm} \\
   \,$z$          & \,0.1105(1)    & 0.11057(3)  & \,0.1102(1)            \rule{0mm}{3.5mm} \\
   \multicolumn{4}{l}{$U_{\mathrm{iso}}$ ($10^{-3}$\AA$^2$)}\rule{0mm}{4.5mm}   \\
   \,Pd           & 5.1(1)         & 5.8(3)      & 1.8(1)                 \rule{0mm}{4mm}   \\
   \,Cr           & 4.4(1)         & 4.6(3)      & 2.3(1)                 \rule{0mm}{3.5mm} \\
   \,O            & 4.4(3)         & 5.3(3)      & 3.7(3)                 \rule{0mm}{3.5mm} \\
  \hline\hline
 \end{tabular*}
 \label{table_1}
\end{center}
\end{table}
\begin{table}
\begin{center}
 \caption{Anisotropic atomic displacement parameters $U_{ij}$ (in units of $10^{-3}$\AA$^2$) of PdCrO$_2$ at RT.
 The parameters were refined with the neutron data from D9.}
 \begin{tabular*}{0.48\textwidth}{@{\extracolsep{\fill}}lcccccc}
  \hline\hline
   \,  Atom & $U_{\mathrm{11}}$ & $U_{\mathrm{22}}$ & $U_{\mathrm{33}}$ & $U_{\mathrm{12}}$ & $U_{\mathrm{13}}$ & $U_{\mathrm{23}}$ \rule{0mm}{3.5mm} \\ \hline
   \,  Pd   & 6.2(3)            & 6.2(3)            & 4.9(3)            & 3.1(3)            & 0                 & 0                 \rule{0mm}{3.5mm} \\ 
   \,  Cr   & 4.1(3)            & 4.1(3)            & 5.5(3)            & 2.0(3)            & 0                 & 0                 \rule{0mm}{3.5mm} \\ 
   \,  O    & 5.2(3)            & 5.2(3)            & 5.4(2)            & 2.6(3)            & 0                 & 0                 \rule{0mm}{3.5mm} \\ 
  \hline\hline
 \end{tabular*}
 \label{table_1.2}
\end{center}
\end{table}

\subsection{Neutron diffraction}
\begin{figure}[t]
\begin{center}
 \includegraphics[width=0.45\textwidth]{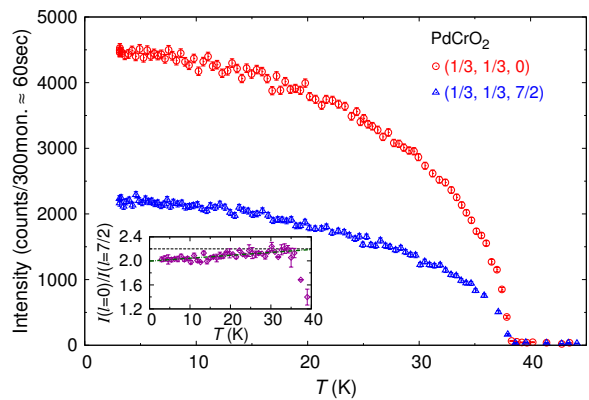}
\caption{
(Color online)
Temperature dependence of the intensity $I(l)$ of magnetic reflections at
(1/3, 1/3, $l$) with $l=0$ and $\frac{7}{2}$.
The inset shows the temperature dependence of 
the intensity ratio $I(l=0)/I(l=\frac{7}{2})$.
The deviation from the constant value for its ratio
below $T_{\rm N}$ suggests that a slight change in the spin configuration appears 
at temperatures below 30~K.
}
\label{fig.1}
\end{center}
\end{figure}

\begin{figure}[t]
\begin{center}
 \includegraphics[width=0.45\textwidth]{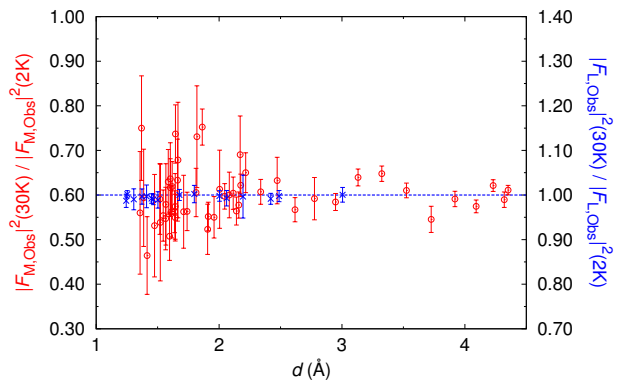}
\caption{
(Color online)
Intensity ratios of magnetic or nuclear reflections taken at 30 and 2~K.
The graph is plotted for the ratios vs the distance between lattice planes, $d$.
The ratio of nuclear reflections ($|F_\mathrm{L,Obs}|^2(30\mathrm{K})/|F_\mathrm{L,Obs}|^2(2\mathrm{K})$) is constant,
while that of magnetic reflections ($|F_\mathrm{M,Obs}|^2(30\mathrm{K})/|F_\mathrm{M,Obs}|^2(2\mathrm{K})$) 
deviates from a constant value.
This result also suggests appearance of a small difference in the magnetic structure 
between 30~K and 2~K.
}
\label{fig.1a}
\end{center}
\end{figure}

Magnetic reflections of PdCrO$_2$ were observed
at $\bm{Q} = (\frac{1}{3}, \frac{1}{3}, l)$ and ($\frac{2}{3}, \frac{2}{3}, l$) with $l=0, \frac{1}{2}, 1, \frac{3}{2}, 2 \cdots$,
being consistent with the previous reports of powder neutron diffraction~\cite{Mekata1995,H.TakatsuPRB2009}.
We confirmed that those magnetic peaks appear at commensurate positions 
within the present experimental accuracies of the 4G and C11 spectrometers.
We did not find any magnetic reflections at 
$\bm{Q}=(00l)$, $(10l)$, $(01l)$,$(11l)$ with $l=0,\frac{1}{2},1,\frac{3}{2},\cdots$.
Figure~\ref{fig.1} shows the temperature dependence of intensities of the magnetic reflections at 
($\frac{1}{3}, \frac{1}{3}, 0$) and ($\frac{1}{3}, \frac{1}{3}, \frac{7}{2}$).
The magnetic peaks appear at temperatures below $T_{\mathrm{N}}$
and their intensities monotonically increase on cooling.
Two successive phase transitions, separated by $\varDelta T = 0.4$~K, 
are observed in the specific heat data~\cite{note_cp_of_PdCrO2}.
These transitions are expected for a small finite $D$ ($<0$)~\cite{Fujiki1983,Blankschtein1984,K.KimuraPRB2008}.
However, such splitting of $T_{\mathrm{N}}$ could not be detected in the neutron diffraction experiment within
the experimental accuracy of 1~K.
This small split of $T_{\mathrm{N}}$ will have to be confirmed by diffraction experiments in future.
The intensity ratio between ($\frac{1}{3}, \frac{1}{3}, 0$) and ($\frac{1}{3}, \frac{1}{3}, \frac{7}{2}$) reflections 
is still slightly temperature dependent below about 20--30~K (the inset of Fig.~\ref{fig.1}).
This feature is also confirmed by intensity ratios of magnetic reflections taken at 2~K and 30~K (Fig.~\ref{fig.1a}).
These results imply that a slight change of magnetic structure occurs around $T^{*}\sim20$~K,
which is in accord with the appearance of UAHE below this temperature.

To analyze the magnetic and crystal structures,
we measured integrated intensities of the Bragg reflections 
at 2~K and 30~K with the four circle spectrometer D10.
%
Observed and calculated squares of the structure factor of nuclear reflections
are listed in Table~I of the Supplemental Material~\cite{supplement_PdCrO2}.
For the calculation, 
we assumed the delafossite structure and refined 
one parameter of the oxygen-ion position $z$.
We obtained $z=0.1104(1)$,
which agrees well with results given in Table~\ref{table_1}.
Due to the secondary extinction effect,
the observed values for the larger intensities tend to deviate from the calculated values,
while those for the smaller intensities are in agreement with calculations (Fig.~\ref{fig.6}).

\subsection{Magnetic structure analysis}
\label{mag.ana}
In order to fit the integrated intensities of the magnetic reflections,
we considered four structure models for the magnetic structure.
All of these consist of spins that lie in a plane containing the hexagonal $c$ axis:
(1) the coplanar single-$q$ 120$^\circ$ spin structure,
where the integer $l$ and half-integer $l$ reflections come from
different $q$ modulation domains [Fig.~\ref{fig.5} (a)];
(2) a single-$q$ structure with a collinear polarization, which also forms multiple domains [Fig.~\ref{fig.5} (b)];
(3) a general multi-$q$ coplanar 120$^\circ$ spin structure, which has a clockwise (+) and anticlockwise (-) rotation 
degree of freedom in each layer [Fig.~\ref{fig.5} (c)];
(4) a non-coplanar spin structure based on the general 120$^\circ$ spin structure of the model (3),
where now the spin plane can rotate around the $c$ axis from one $z$-layer to the next [Fig.~\ref{fig.5} (d)].
This rotational misfit is characterized by the normal vector to the 120$^\circ$-spin plane,
which is considered to point into different directions for each $z$-layer.
The orientation of the 120$^\circ$-spin plane can be described by different azimuthal angles of its normal vector 
$\alpha_n$ ($n=0,1,2,3,4,5$) for each layer.
We performed a least-squares fit by these models
and found that models (3) and (4) provide solutions that account for the observed intensities.
The representation analysis revealed that these magnetic structures can be classified by 
using small representations deduced from the $R\bar{3}m$ crystal symmetry (Sect.~\ref{sec.rep} and Appendix).
The details of analysis are as follows.

The intensity of the magnetic reflection at a wave vector $\bm{Q}$ is written as 
\begin{alignat}{2}
&I = (\gamma r_0)^2 |F_{\mathrm{M}}(\bm{Q})|^2 \Bigl[\frac{g}{2} f(\bm{Q}) \Bigr]^2, \\
&F_{\mathrm{M}}(\bm{Q}) = \sum_{\substack{ \mathrm{\ magnetic} \\ \mathrm{unit\, cell} }} \exp({i\bm{Q}\cdot\bm{R}}) 
                     \bigr[ \bm{S}_{\bm{R}} - \hat{\bm{Q}}(\hat{\bm{Q}}\cdot\bm{S}_{\bm{R}}) \bigl],
\label{eq.2}
\end{alignat}
where $\gamma=1.913$, $r_0$ is the classical radius of the electron,
$g$ and $f(\bm{Q})$ are the $g$ factor and magnetic form factor of Cr$^{3+}$, respectively.
Here, we assumed $g\simeq2$ following the result of the electron spin resonance spectroscopy~\cite{Hemmeida2011}.
$\hat{\bm{Q}}$ is the unit vector along the wave-vector transfer $\bm{Q}$.
In Eq.~(2), the temperature factor is neglected.
For the analysis,
we have taken an average of $|F_{\mathrm{M}}(\bm{Q})|^2$ over magnetic structure domains,
which are naturally derived by symmetry operations of the space group $R\bar{3}m$.

Magnetic structure models of PdCrO$_2$ that we consider consist of 18 sublattice structures, whose
18 magnetic Cr sites A$_n$, B$_n$, and C$_n$ ($n=0,1,2,3,4,5$) are shown in Fig.~\ref{fig.5}(e). 
The hexagonal coordinates of the sublattice sites $\bm{R}_{\text{A}_n}$, $\bm{R}_{\text{B}_n}$, 
and $\bm{R}_{\text{C}_n}$ are 
\begin{alignat}{4}
\bm{R}_{\text{A}_0} &= \left(          0,           0, \frac{1}{2} \right), 
\bm{R}_{\text{A}_1}  = \left(\frac{2}{3}, \frac{1}{3}, \frac{5}{6} \right), 
\bm{R}_{\text{A}_2}  = \left(\frac{4}{3}, \frac{2}{3}, \frac{7}{6} \right) \notag \\
\bm{R}_{\text{A}_3} &= \left(          0,           0, \frac{3}{2} \right), 
\bm{R}_{\text{A}_4}  = \left(\frac{2}{3}, \frac{1}{3}, \frac{11}{6} \right), 
\bm{R}_{\text{A}_5}  = \left(\frac{4}{3}, \frac{2}{3}, \frac{13}{6} \right) \notag \\
\bm{R}_{\text{B}_n} &= \bm{R}_{\text{A}_n} + (1,0,0), \notag \\
\bm{R}_{\text{C}_n} &= \bm{R}_{\text{A}_n} + (2,0,0). \notag
\end{alignat}

The observation of no magnetic intensity at reflections 
$\bm{Q}=(00l)$, $(10l)$, $(01l)$, $(11l)$ with $l=0, \frac{1}{2}, 1, \frac{3}{2}, \cdots$
indicates that the 18 sublattice spins $\bm{S}_{\text{A}_n}$, $\bm{S}_{\text{B}_n}$, and $\bm{S}_{\text{C}_n}$  
satisfy constraints
\begin{equation}
\bm{S}_{\text{A}_n} + \bm{S}_{\text{B}_n} + \bm{S}_{\text{C}_n} = 0.
\label{eq.4} 
\end{equation}
These constraints are alternatively expressed by a six-$q$ structure
\begin{align}
\bm{S}_{\text{X}_n} &= \sum_{j=1}^{6} \left( \bm{a}_j \exp[i \bm{q}_j \cdot \bm{R}_{\text{X}_n}] + \bm{a}_j^{\ast} \exp[- i \bm{q}_j \cdot \bm{R}_{\text{X}_n}] \right) \notag \\
&= \sum_{j=1}^{6} 2 \left( \bm{a}_j^{\prime} \cos[\bm{q}_j \cdot \bm{R}_{\text{X}_n}] + \bm{a}_j^{\prime \prime} \sin[\bm{q}_j \cdot \bm{R}_{\text{X}_n}] \right), 
\label{eq.magst}
\end{align}
where $\text{X}_n$ stands for A$_n$, B$_n$, or C$_n$,
and wave vectors $\bm{q}_j$ are
\begin{align}
\bm{q}_1 &= \left(  \frac{1}{3},  \frac{1}{3}, 0 \right), 
\bm{q}_2  = \left( -\frac{2}{3},  \frac{1}{3}, 0 \right), 
\bm{q}_3  = \left(  \frac{1}{3}, -\frac{2}{3}, 0 \right), \notag \\
\bm{q}_4 &= \left(  \frac{1}{3},  \frac{1}{3}, \frac{1}{2} \right), 
\bm{q}_5  = \left( -\frac{2}{3},  \frac{1}{3}, \frac{1}{2} \right), 
\bm{q}_6  = \left(  \frac{1}{3}, -\frac{2}{3}, \frac{1}{2} \right).
\end{align}
In Eq.~(\ref{eq.magst}), 
$\bm{a}_j = \bm{a}_j^{\prime} - i \bm{a}_j^{\prime \prime}$ are complex vectors, 
whereas $\bm{a}_j^{\prime}$ and $\bm{a}_j^{\prime \prime}$ are real vectors. 

In principle there are 36 adjustable parameters for the present magnetic structure
determination (12 real or 6 complex vectors),
we therefore considered the simplified structure models (1)--(4) to reduce number of fitting parameters.
Note that as discussed later,
we confirmed that all of these magnetic structures were classified by using symmetry properties 
of the space group $R\bar{3}m$ of the crystal symmetry.

\subsubsection{Model (1): single-$q$ 120$^\circ$ spin structure}
We assume multiple domains of a single-$q$ structure, 
which consists of a 120$^\circ$ spin plane including the $c$ axis
[Fig.~\ref{fig.5}(a)]. 
This is the simplest structure model deduced from the spin Hamiltonian 
of Eq.~(\ref{eq.0}).
The magnetic structure of one domain responsible for integer-$l$ reflections is  
\begin{align}
\bm{S}_{\text{X}_n} &= 
  S\hat{{\bm z}}         \cos[ \phi + \bm{q}_1 \cdot \bm{R}_{\text{X}_n}] 
+ S\hat{\bm{e}}_{\alpha} \sin[ \phi + \bm{q}_1 \cdot \bm{R}_{\text{X}_n}], 
\label{eq.singleq1} \\
\hat{\bm{e}}_{\alpha} &= \hat{{\bm x}} \cos \alpha + \hat{{\bm y}} \sin \alpha, \notag
\end{align}
where $\hat{{\bm x}}$, $\hat{{\bm y}}$, and $\hat{{\bm z}}$ are orthogonal unit vectors
($\hat{{\bm a}} = a \hat{{\bm x}}$, $\hat{{\bm b}} = a (-\frac{1}{2}\hat{{\bm x}} + \frac{\sqrt{3}}{2}\hat{{\bm y}} )$,
and $\hat{{\bm c}} = c\hat{{\bm z}}$),
$\alpha$ and $\phi$ are constants.
Note that Eq.~(\ref{eq.singleq1}) is deduced from Eq.~(\ref{eq.magst}) by substituting 
$\bm{a}_1 = \frac{S}{2} (\hat{{\bm z}} - i \hat{\bm{e}}_{\alpha}) e^{i\phi}$.
Each spin of this domain can be written as 
\begin{align}
\bm{S}_{\text{A}_n} &= 
  S \left[ \hat{ \bm{z}} \cos \phi_{1,n} + 
         \hat{\bm{e}}_{\alpha} \sin \phi_{1,n}   \right], \notag \\
\bm{S}_{\text{B}_n} &= 
  S \left[ \hat{ \bm{z}} \cos \left( \phi_{1,n} + \frac{2 \pi}{3} \right)
+        \hat{\bm{e}}_{\alpha}  \sin \left( \phi_{1,n}  + \frac{2 \pi}{3} \right) \right], \notag \\
\bm{S}_{\text{C}_n} &= 
  S \left[ \hat{ \bm{z}} \cos \left( \phi_{1,n} - \frac{2 \pi}{3} \right)
+        \hat{\bm{e}}_{\alpha}  \sin \left( \phi_{1,n} - \frac{2 \pi}{3} \right) \right], 
\end{align}
where $\phi_{1,n} = \phi + \frac{2\pi}{3}n$,
and $\phi$ is assumed to be zero in the analysis.
Symmetrically equivalent domains responsible for integer-$l$ reflections
are obtained by transformations of 
the space group operations with respect to $\bm{q}_1 \rightarrow \bm{q}_j$ ($j = 2, 3$). 

One magnetic domain responsible for half-integer-$l$ reflections consists of 
the spin vectors
\begin{align}
\bm{S}_{\text{X}_n} &= 
  S\hat{{\bm z}}         \cos[ \phi + \bm{q}_4 \cdot \bm{R}_{\text{X}_n}] 
+ S\hat{\bm{e}}_{\alpha} \sin[ \phi + \bm{q}_4 \cdot \bm{R}_{\text{X}_n}],  
\label{eq.singleq2} \\
\hat{\bm{e}}_{\alpha} &= \hat{{\bm x}} \cos \alpha + \hat{{\bm y}} \sin \alpha, \notag
\end{align}
where $\alpha$ and $\phi$ are constants.
Here, Eq.~(\ref{eq.singleq2}) is equivalent to Eq.~(\ref{eq.magst}) with 
$\bm{a}_4 = \frac{S}{2} (\hat{{\bm z}} - i \hat{\bm{e}}_{\alpha}) e^{i\phi}$,
as in the case of the integer-$l$ domain.
Each spin is 
\begin{align}
\bm{S}_{\text{A}_n} &= 
  S \left[ \hat{ \bm{z}} \cos \phi_{4,n} + 
         \hat{\bm{e}}_{\alpha} \sin \phi_{4,n}   \right], \notag \\
\bm{S}_{\text{B}_n} &= 
  S \left[ \hat{ \bm{z}} \cos \left( \phi_{4,n} + \frac{2 \pi}{3} \right)
+        \hat{\bm{e}}_{\alpha}  \sin \left( \phi_{4,n}  + \frac{2 \pi}{3} \right) \right], \notag \\
\bm{S}_{\text{C}_n} &= 
  S \left[ \hat{ \bm{z}} \cos \left( \phi_{4,n} - \frac{2 \pi}{3} \right)
+        \hat{\bm{e}}_{\alpha}  \sin \left( \phi_{4,n} - \frac{2 \pi}{3} \right) \right], 
\end{align}
where $\phi_{4,n} = \phi + \sigma_n \frac{\pi}{2}$
and $\phi$ is assumed to be zero.
Here, $\sigma_n=(-1)^n$.
Symmetrically equivalent domains are obtained by transformations of 
the space group operations with respect to 
$\bm{q}_4 \rightarrow \bm{q}_j$ ($j = 5, 6$). 

\subsubsection{Model (2): single-$q$ collinear spin structure}
We assume a single-$q$ collinear structure with multiple domains [Fig.~\ref{fig.5}(b)]. 
The magnetic structure of a domain, which is responsible for integer-$l$ reflections,
is described by e.g.,
\begin{align}
\bm{S}_{\text{X}_n} &= \bm{a}_1 \exp[i \bm{q}_1 \cdot \bm{R}_{\text{X}_n}] 
                     + \bm{a}_1^{\ast} \exp[- i \bm{q}_1 \cdot \bm{R}_{\text{X}_n}] 
\label{eq.single_q_co_1}\\
\bm{a}_1   &= \frac{1}{2} S (\cos\beta\hat{{\bm z}} + \sin\beta\hat{\bm{e}}_{\alpha}) \exp(i\phi), \notag  
\end{align}
where $S$, $\beta$, and $\phi$ are constants.
Each spin is
\begin{align}
&\bm{S}_{\text{A}_n} = 
 S (\cos\beta\hat{{\bm z}} + \sin\beta\hat{\bm{e}}_{\alpha}) \cos \phi_{1,n}, \notag \\
&\bm{S}_{\text{B}_n} = 
 S (\cos\beta\hat{{\bm z}} + \sin\beta\hat{\bm{e}}_{\alpha}) \cos \left( \phi_{1,n} + \frac{2 \pi}{3} \right), \notag \\
&\bm{S}_{\text{C}_n} = 
 S (\cos\beta\hat{{\bm z}} + \sin\beta\hat{\bm{e}}_{\alpha}) \cos \left( \phi_{1,n} - \frac{2 \pi}{3} \right), \notag \\
&\hat{\bm{e}}_{\alpha} = \hat{{\bm x}} \cos \alpha + \hat{{\bm y}} \sin \alpha,  \notag \\
\end{align}
where $\phi_{1,n} = \phi + \frac{2\pi}{3}n$ ($n=0,1,2,3,4,5$)
and $\phi$ is assumed to be zero.
Symmetrically equivalent domains are obtained by transformations of 
the space group operations with respect to $\bm{q}_1 \rightarrow \bm{q}_j$ ($j = 2, 3$).

A magnetic domain, which provides half-integer-$l$ reflections,
is 
\begin{align}
\bm{S}_{\text{X}_n} &= \bm{a}_4 \exp[i \bm{q}_4 \cdot \bm{R}_{\text{X}_n}] 
                     + \bm{a}_4^{\ast} \exp[- i \bm{q}_4 \cdot \bm{R}_{\text{X}_n}] \
\label{eq.single_q_co_2}\\
\bm{a}_4   &= \frac{1}{2} S (\cos\beta\hat{{\bm z}} + \sin\beta\hat{\bm{e}}_{\alpha}) \exp(i\phi). \notag  
\end{align}
Each spin is written by
\begin{align}
&\bm{S}_{\text{A}_n} = 
  S (\cos\beta\hat{{\bm z}} + \sin\beta\hat{\bm{e}}_{\alpha}) \cos \phi_{4,n}, \notag \\
&\bm{S}_{\text{B}_n} = 
  S (\cos\beta\hat{{\bm z}} + \sin\beta\hat{\bm{e}}_{\alpha}) \cos \left( \phi_{4,n} + \frac{2 \pi}{3} \right), \notag \\
&\bm{S}_{\text{C}_n} = 
  S (\cos\beta\hat{{\bm z}} + \sin\beta\hat{\bm{e}}_{\alpha}) \cos \left( \phi_{4,n} - \frac{2 \pi}{3} \right), \notag \\
&\hat{\bm{e}}_{\alpha} = \hat{{\bm x}} \cos \alpha + \hat{{\bm y}} \sin \alpha, \notag \\
\end{align}
where $\phi_{4,n} = \phi + \sigma_n \frac{\pi}{2}$ and 
$\phi$ is assumed to be zero.
Here, $\sigma_n=(-1)^n$.
Symmetrically equivalent domains are obtained by transformations of 
the space group operations with respect to $\bm{q}_4 \rightarrow \bm{q}_j$ ($j = 5, 6$).

\subsubsection{Model (3): coplanar 120$^\circ$ spin structure}
For the model (3), 
we consider as before a coplanar 120$^\circ$ spin structure in a plane parallel to the $c$ axis [Fig.~\ref{fig.5}(c)].
Each spin is now written as
\begin{align}
&\bm{S}_{\text{A}_n} = 
  S \left[ \hat{ \bm{z}} \cos \phi_n + 
         \hat{\bm{e}}_{\alpha} \sin \phi_n   \right], \notag \\
&\bm{S}_{\text{B}_n} = 
  S \left[ \hat{ \bm{z}} \cos \left( \phi_n + \xi_n \frac{2 \pi}{3} \right)
+        \hat{\bm{e}}_{\alpha} \sin \left( \phi_n + \xi_n \frac{2 \pi}{3} \right) \right], \notag \\
&\bm{S}_{\text{C}_n} = 
  S \left[ \hat{ \bm{z}} \cos \left( \phi_n - \xi_n \frac{2 \pi}{3} \right)
+        \hat{\bm{e}}_{\alpha} \sin \left( \phi_n - \xi_n \frac{2 \pi}{3} \right) \right], \notag \\
&\hat{\bm{e}}_{\alpha} = 
\hat{{\bm x}} \cos \alpha + \hat{{\bm y}} \sin \alpha, \notag \\
\end{align}
where $\phi_n$ ($n=0,1,2,3,4,5$) are constants, 
which represent rotation angles of the 120$^\circ$ spin 
around the normal vector of the plane consisting of the $\hat{\bm{c}}$ and $\hat{\bm{e}}_{\alpha}$ vectors.
Here we have introduced an additional degree of freedom $\xi_n = \pm 1$, 
which describes the rotation direction of spins:
$\xi_n = +1$ represents the clockwise rotation, 
while $\xi_n = -1$ describes
the anticlockwise rotation.
This parameter $\xi_n$ allows to test 32 different structures which can be grouped into
eight independent classes of spin structures, related to the rotation sense of spins
in each layer, that is,
$++++++$, $+++++-$, $++++--$, $+++-+-$,
$++-++-$, $+++---$, $++-+--$, $+-+-+-$.
Here, 
$+$, $-$ stand for $\xi_n = +1$, $-1$, respectively.

\subsubsection{Model (4): non-coplanar 120$^\circ$ spin structure}
For the model (4), 
we consider a non-coplanar 120$^\circ$ spin structure where 
the non-coplanarity of spin configurations is introduced by different orientations of
the 120$^\circ$ spin-planes in subsequent $z$-layers [Fig.~\ref{fig.5}(d)].
{\color[rgb]{0,0,0} 
Spin vectors of this structure are 
\begin{align}
&\bm{S}_{\text{A}_n} = 
  S \Bigl[ \hat{ \bm{z}} \cos \gamma_n \cos \phi_n + 
           \hat{\bm{e}}_{\alpha_{n}} \sin \phi_n + 
           \hat{\bm{e}}_{\alpha_{n}'} \sin \gamma_n \cos \phi_n  \Bigr], \notag \\
&\bm{S}_{\text{B}_n} = 
  S \biggl[ \hat{ \bm{z}} \cos \gamma_n \cos \left( \phi_n + \xi_n \frac{2 \pi}{3} \right)\notag\\
&\hspace{5pt} +  \hat{\bm{e}}_{\alpha_{n}} \sin \left( \phi_n + \xi_n \frac{2 \pi}{3} \right) 
+ \hat{\bm{e}}_{\alpha_{n}'} \sin \gamma_n \cos \left( \phi_n + \xi_n \frac{2 \pi}{3} \right) \biggr], \notag \\
&\bm{S}_{\text{C}_n} = 
  S \biggl[ \hat{ \bm{z}} \cos \gamma_n \cos \left( \phi_n - \xi_n \frac{2 \pi}{3} \right) \notag\\
&\hspace{5pt} +  \hat{\bm{e}}_{\alpha_{n}} \sin \left( \phi_n - \xi_n \frac{2 \pi}{3} \right)
+ \hat{\bm{e}}_{\alpha_{n}'} \sin \gamma_n \cos \left( \phi_n - \xi_n \frac{2 \pi}{3} \right) \biggr], \notag \\
\notag\\
&\hat{\bm{e}}_{\alpha_{n}} = 
\hat{{\bm x}} \cos \alpha_{n} + \hat{{\bm y}} \sin \alpha_{n}, \notag \\
&\hat{\bm{e}}_{\alpha_{n}'} = 
\hat{{\bm x}} \cos \Bigl(\alpha_{n}+\frac{\pi}{2}\Bigr) + \hat{{\bm y}} \sin \Bigl(\alpha_{n}+\frac{\pi}{2}\Bigr), 
\label{eq.model4}
\end{align}
where $\gamma_{n}$, $\alpha_n$ and $\phi_n$ ($n=0,1,2,3,4,5$) are constants,
indicating polar angles from the $z$ axis to 
the normal vector of the plane consisting of the $\hat{\bm{c}}$ and $\hat{\bm{e}}_{\alpha}$ vectors,
azimuthal angles of the spin-plane normal in the different $z$-layers,
and rotation angles of the 120$^\circ$ spin around the normal vector, respectively.
In the analysis, %
we assume that in each $z$-layer the 120$^\circ$ plane is parallel to the $c$ axis   
(i.e., $\gamma_{n}=0$). 
We also consider a case that there is a misfit between 
$\alpha_n$ of even ($n=0,2,4$) layers and that of odd ($n=1,3,5$) layers. 
This structure is, however, represented by irreducible representations
of the $R\bar{3}m$ crystal symmetry, which will be discussed in detail in Sect.~\ref{sec.rep}. 
In Eq.~(\ref{eq.model4}), the definition of $\xi_n$ is the same as that of the model~(3).
Here, we fixed $\xi_n$ 
so that $\xi_n=+1$ ($n=0,2,4$) and $\xi_n=-1$ ($n=1,3,5$). 
Note that we can consider other combinations in these parameters,
however they do not change the general fitting result because 
misfit angles $\alpha_n$ between layers are fitting parameters,
namely certain values of $\alpha_n$ can represent $\xi_1 = \pm1$;
e.g., 
$\alpha_1 = 0^\circ$, $\alpha_2 = 180^\circ$ and $\xi_1 = \xi_2 = +1$
represent the same structure as $\xi_1 = +1$ and $\xi_2 = -1$ for $\alpha_1 = \alpha_2 = 0^\circ$
(in these, a case of $\phi_1 = \phi_2 =0^\circ$ is considered).
}

\subsubsection{Fit results of the model structures}
\label{sec.eva}
\begin{table}
\begin{center}
 \caption{Minimum values of $\chi^2$, $R$ factors, and magnetic moments for models (1)--(4).
Values of the model (3) are based on one of the least-square solutions of 
the $+-+-+-$ structure which is the best fitted structure in the model (3). 
Large $\chi^2$ value of the model (1) and magnetic moment of the model (2) beyond the expected value
of 3~$\mu_\mathrm{B}$ indicate that these models can be ruled out for the magnetic structure of PdCrO$_2$.}
 \begin{tabular*}{0.48\textwidth}{@{\extracolsep{\fill}}lllll}
  \hline\hline
                       & model(1)                     & model(2)                     & model(3)             & model(4)             \rule{0mm}{3.5mm} \\ \hline
    $\chi^2$(2~K)      & 2376                         & 50                           & 57                   & 56                   \rule{0mm}{3.5mm} \\ 
    $R_{\rm wp}$(2~K)  & 46.27\%                      & 6.69\%                       & 7.17\%               & 7.08\%               \rule{0mm}{3.5mm} \\ 
    $R_{\rm e}$(2~K)   & 6.97\%                       & 6.84\%                       & 7.35\%               & 7.35\%               \rule{0mm}{3.5mm} \\ 
    $R_{\rm B} $(2~K)  & 37.68\%                      & 5.20\%                       & 5.80\%               & 5.68\%               \rule{0mm}{3.5mm} \\ 
    $\chi^2$(30~K)     & 1707                         & 51                           & 74                   & 60                   \rule{0mm}{6.5mm} \\ 
    $R_{\rm wp}$(30~K) & 45.03\%                      & 7.76\%                       & 9.40\%               & 8.41\%               \rule{0mm}{3.5mm} \\ 
    $R_{\rm e}$(30~K)  & 8.01\%                       & 7.86\%                       & 8.44\%               & 8.44\%               \rule{0mm}{3.5mm} \\ 
    $R_{\rm B} $(30~K) & 37.47\%                      & 6.88\%                       & 8.10\%               & 7.22\%               \rule{0mm}{3.5mm} \\ 
    Magnetic           & 1.82$\mu_\mathrm{B}$         & 3.15$\mu_\mathrm{B}$         & 2.20$\mu_\mathrm{B}$ & 2.20$\mu_\mathrm{B}$ \rule{0mm}{6.5mm} \\ 
    moment (2K)        & {\scriptsize(integer $l$)}   & {\scriptsize(integer $l$)}   &                      &                      \rule{0mm}{2.5mm} \\ 
                       & 1.98$\mu_\mathrm{B}$         & 3.08$\mu_\mathrm{B}$         &                      &                      \rule{0mm}{3.5mm} \\ 
                       & {\scriptsize(half-int. $l$)} & {\scriptsize(half-int. $l$)} &                      &                      \rule{0mm}{2.5mm} \\ 
  \hline\hline
 \end{tabular*}
 \label{table.3}
\end{center}
\end{table}
\begin{table}
\begin{center}
 \caption{Minimum value of $\chi^2$ for one of the least-square solutions of the model (3)
 in all eight independent spin structure classes.}
 \begin{tabular*}{0.48\textwidth}{@{\extracolsep{\fill}}lll}
  \hline\hline
   \,            & 2K    & 30K \quad\qquad         \rule{0mm}{3.5mm} \\ \hline
   \,  $++++++$  & 2372  & 1710\quad\qquad         \rule{0mm}{3.5mm} \\ 
   \,  $+++++-$  & 901   & 634 \quad\qquad         \rule{0mm}{3.5mm} \\ 
   \,  $++++--$  & 717   & 546 \quad\qquad         \rule{0mm}{3.5mm} \\ 
   \,  $+++-+-$  & 227   & 171 \quad\qquad         \rule{0mm}{3.5mm} \\ 
   \,  $+++---$  & 448   & 388 \quad\qquad         \rule{0mm}{3.5mm} \\ 
   \,  $++-+--$  & 961   & 548 \quad\qquad         \rule{0mm}{3.5mm} \\ 
   \,  $+-+-+-$  & 57    & 74  \quad\qquad         \rule{0mm}{3.5mm} \\ 
  \hline\hline
 \end{tabular*}
 \label{table.3.2}
\end{center}
\end{table}
\begin{figure}[t]
\begin{center}
 \includegraphics[width=0.45\textwidth]{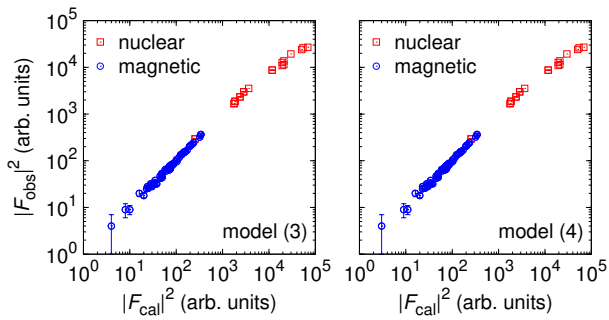}
\caption{
(Color online)
Observed and calculated values of the squared nuclear and magnetic structure factors. 
The results at 2~K are presented for 
the $+-+-+-$ 120$^\circ$ coplanar spin structure in a plane including the $c$ axis 
[model (3)] and for the non-coplanar spin structure [model (4)].
These models are considered meaningful for the magnetic structure of PdCrO$_2$.
}
\label{fig.6}
\end{center}
\end{figure}
\begin{figure*}[t]
\begin{center}
 \includegraphics[width=0.95\textwidth]{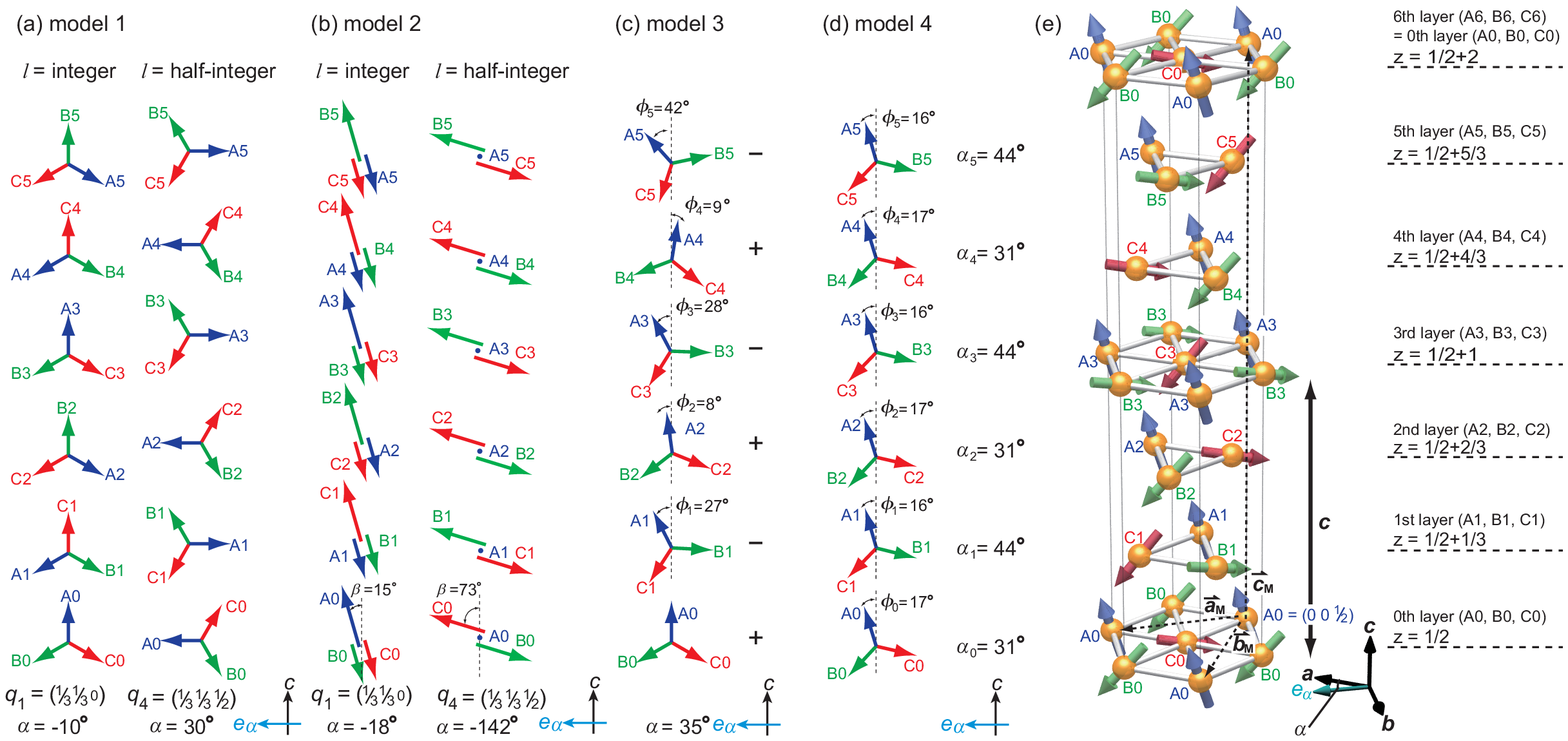}
\caption{
(Color online)
Magnetic structure models of PdCrO$_2$.
(a) One solution of a single-$q$ 120$^\circ$ spin structure of the model (1) at 2~K.
Here, the solution of the domain given by $\bm{q}_1$ and $\bm{q}_4$.
The magnetic domain of $\bm{q}_1 = (\frac{1}{3}, \frac{1}{3}, 0)$ gives rise to 
integer-$l$ reflections of ($\frac{1}{3}, \frac{1}{3}, l$), while 
that of $\bm{q}_4 = (\frac{1}{3}, \frac{1}{3}, \frac{1}{2}$) provide half-integer-$l$ reflections.
18 arrows labeled by A$_n$, B$_n$ and C$_n$ ($n=0,1,...5$)
represent 18 sublattice spins $S_{\mathrm{A}_n}$, $S_{\mathrm{B}_n}$ and $S_{\mathrm{C}_n}$, 
respectively.
(b) A single-$q$ collinear spin structure of the model (2). 
Magnetic domains of $\bm{q}_1$ and $\bm{q}_4$ are shown.
Closed circles in the structure of $\bm{q}_4$ indicate $S=0$.
(c) One of the solutions of the coplanar 120$^\circ$ spin structure of the model (3) at 2~K. 
Clockwise and anticlockwise rotational directions are represented by $+$ and $-$. 
The alternative stacks of the $+$ and $-$ layers shown in the figure represent
the best fit result, i.e., the $+-+-+-$ structure.
(d) One of the solutions of the non-coplanar spin structure of the model (4) at 2~K.
The non-coplanarity of spins arises from a rotation of the 120$^\circ$ spin plane around 
the $c$ axis in subsequent $z$-layers, that is, 
the azimuthal angles of the normal vector to the 120$^\circ$ spin plane, 
$\alpha_n$ ($n=0,1,2,3,4,5$), are different for each layer.
Here a case of which there is only a difference between $\alpha_n$ of even layers ($n=0,2,4$) and 
that of odd layers ($n=1,3,5$) is considered.
(e) Schematic drawing of layered triangular network of Cr atoms with spins of the model structure of (d).
$\hat{{\bm a}}_\mathrm{M}$, $\hat{{\bm b}}_\mathrm{M}$ and $\hat{{\bm c}}_\mathrm{M}$ represent the magnetic unit cell vectors,
consisting of the magnetic structure of the 18 sublattices of Cr ions:
$\hat{{\bm a}}_\mathrm{M} = 2\hat{{\bm a}} + \hat{{\bm b}}$, $\hat{{\bm b}}_\mathrm{M} = \hat{{\bm a}} + 2\hat{{\bm b}}$, 
and $\hat{{\bm c}}_\mathrm{M} = 2 \hat{{\bm c}}$.
Here, $\hat{{\bm a}}$, $\hat{{\bm b}}$, and $\hat{{\bm c}}$ are the unit-cell vectors 
for the hexagonal lattice setting 
($\hat{{\bm a}} = a \hat{{\bm x}}$, $\hat{{\bm b}} = a (-\frac{1}{2}\hat{{\bm x}} + \frac{\sqrt{3}}{2}\hat{{\bm y}} )$,
$\hat{{\bm c}} = c\hat{{\bm z}}$, where $\hat{{\bm x}}$, $\hat{{\bm y}}$, and $\hat{{\bm z}}$ are
orthogonal unit vectors, and $a$ and $c$ are the lattice constants, respectively).
}
\label{fig.5}
\end{center}
\end{figure*}

Least square fits were performed with four parameters for the model (1);
$S$ and $\alpha$ for both integer- and half-integer-$l$ domain structures.
For the model (2), 
we used six parameters;
$S_{x}$, $S_{y}$, $S_{z}$ for both integer- and half-integer-$l$ domain structures.
Here, $S_{x}$, $S_{y}$ and $S_{z}$ are parameters having a relation that
$\bm{a}_j  = \frac{1}{2}S \left( \cos\beta\hat{{\bm z}} + \sin\beta\hat{\bm{e}}_{\alpha} \right)e^{i\phi}
           = (S_{x}\hat{{\bm x}} + S_{y}\hat{{\bm y}} + S_{z}\hat{{\bm z}})e^{i\phi}$.
The model (3) considered seven parameters,
$S$, $\alpha$, $\phi_{n}$ ($n=1,2,3,4,5$),
and examined eight independent classes of magnetic structures.
In the analysis, 
we fixed $\phi_0=0^{\circ}$ and considered deviation from it for other angles of $\phi_{n}$.
For the model (4),
we considered five parameters,
$S$, $\phi_i$, $\phi_j$, $\alpha_i$, and $\alpha_j$ ($i=0,2,4$, $j=1,3,5$).
The fit results are summarized in Tables~II, III, IV, and V in 
the Supplemental Material~\cite{supplement_PdCrO2}.

Spin structures corresponding to a set of obtained parameters at 2~K are shown in Figs.~\ref{fig.5}(a)-(d).
$\chi^2$ values of each result and $R$-factors are summarized in Tables~\ref{table.3} and \ref{table.3.2}.
Here, $\chi^2$ is defined by
\begin{alignat}{1}
\chi^2 = \sum_{i=1}^m \biggr (\frac{|F_\mathrm{M}|^2_{\mathrm{obs},i} - |F_\mathrm{M}|^2_{\mathrm{cal},i}}{\sigma(|F_\mathrm{M}|^2_{\mathrm{obs},i})} \biggl)^2,
\label{eq.17}
\end{alignat}
where $m=58$ is the number of the observed magnetic reflections.
A good fit requires $S' = [\chi^2/(m-\delta)]^{\frac{1}{2}} \lessapprox 1.3$ where 
the number of fit parameters $\delta$ is subtracted from the number of reflections $m$.
For the model (1),
we found  $S'(2\mathrm{K}) = 6.6$ and $S'(30\mathrm{K}) = 5.7$.
Fits with the model (1) and spins in the $ab$ plane did not improve the value of $S'$;
$S'(2~\mathrm{K}) = 8.2$ and $S'(30~\mathrm{K}) = 7.3$.
For the model (2),
we found good fit results, 
$S'(2\mathrm{K}) = 1.0$ and $S'(30\mathrm{K}) = 1.0$, but an unphysically large ordered moment 
as discussed below (cf. Table~\ref{table.3}).
Within the model (3),
the $+-+-+-$ structure class yields by far the best fit result (Table.~\ref{table.3.2}),
with $S'(2\mathrm{K}) = 1.1$ and $S'(30\mathrm{K}) = 1.2$.
We obtained several solutions in this structure class with the same values of $S'$.
When evaluating the explicit mathematical form of Eq.~(\ref{eq.magst}),
we always find two large near-equal amplitudes, either $|\bm{a}_{1}|$ and $|\bm{a}_{5}|$ for ($\bm{q}_1, \bm{q}_5$) 
or, equivalently, $|\bm{a}_{2}|$ and $|\bm{a}_{6}|$ for  ($\bm{q}_2, \bm{q}_6$), 
or $|\bm{a}_{3}|$ and $|\bm{a}_{4}|$ for ($\bm{q}_3, \bm{q}_4$)
with all other amplitudes being an order of magnitude smaller.
This resembles the case of LiCrO$_2$ which is an analogous magnet with the same arrangement of
Cr-sites, and implies a double-$q$ structure~\cite{Kadowaki1995}.
We will discuss this result on the basis of the representation analysis 
in the later section.
For the model (4),
an acceptable $S'$ value was obtained
for a case that $\varDelta\alpha_n = \alpha_n - \alpha_{n-1} \leq40^\circ$:
$S'(2\mathrm{K}) = 1.1$ and $S'(30\mathrm{K}) = 1.1$.
Within this model we also obtained several solutions, all of which indicate a finite $\varDelta\alpha_n$,
leading to a non-coplanar spin configuration.
Similar to the case of the model(3), two amplitudes $|\bm{a}_j|$ are always large,
the other small.
The estimated value of the average local scalar spin chirality 
given in Eq.~(\ref{eq.10}) is almost the same for all solutions,
although there are small differences in the value of $\phi_n$ and that of $\alpha_n$ among the solutions.

The estimated magnetic moments at 2~K are listed in Table~\ref{table.3}.
The magnetic moment value for the model (2) clearly exceeds 
the expected value of 3$\mu_\mathrm{B}$ for the Cr$^{3+}$ magnetic system~\cite{P.J.Brown2002,Kadowaki1990},
which is unphysical. We therefore can rule out the model (2).
Instead, the values for models (3) and (4) are about 30\% smaller than the expectation.
For 2D spin systems,
the magnetic moment is indeed expected to be smaller than $g\mu_{\mathrm{B}} S$ due to quantum effects~\cite{Mekata1993,Kadowaki1995}.
Moreover,
a reduction of the magnetic moment is known in materials with
antiferromagnetic coupling and covalent bonding~\cite{P.J.Brown2002}.
It can be attributed to hybridization of orbitals between magnetic and neighboring-ligand ions.
Therefore,
the result of models (3) and (4) can be considered relevant. 
We note that the magnetic moment is the same
in all the solutions in both models (3) and (4).
With these results and the requirement for the $\chi^2$ value, 
at 2~K
the structure of the $+-+-+-$ structure of the model~(3) or 
the model~(4) are considered to be plausible for the magnetic structure of PdCrO$_2$. 
At 30~K, 
the analysis provided almost the same result as at 2~K, with a slight preference of the model~(4)
compared to the model~(3), cf. Table~\ref{table.3.2}.
In Sect.~\ref{sec.dis},
we discuss these magnetic structures in the context of the UAHE.

{\color[rgb]{0,0,0} 
\subsubsection{Representation analysis}
\label{sec.rep}
Before going to the discussion paragraph of Sect.~\ref{sec.dis},
we here discuss symmetry properties of the magnetic structures
from the representation analyses. 
We focus on the analysis using model magnetic structures deduced from
the irreducible representation of the crystal symmetry,
and then discuss symmetries of models~(3) and (4).
Characterizations of other models (1) and (2) and details of the symmetry argument are summarized in the appendix.

From fit results of models (3) and (4) discussed in the former section (Sect.~\ref{sec.eva}),
let us consider a double-$q$ structure such that consists of ($\bm{q}_1, \bm{q}_5$), ($\bm{q}_2, \bm{q}_6$), or ($\bm{q}_3, \bm{q}_4$).
In view of the representation analysis,
these combinations are the simplest combinations that can lead to a magnetic structure having a 120$^\circ$-spin plane,
since the basis vectors of such two wave vectors consist of similar components (Table~\ref{table_A3}).
The double-$q$ structure is written as, for example,
\begin{alignat}{2}
&\bm{S}(\bm{R}) = \sum_{j=3}^{4} \left\{ \bm{b}_j \exp(i\bm{q}_{j}\cdot\bm{R}) + \bm{b}_j^* \exp(-i\bm{q}_{j}\cdot\bm{R})\right\},
\label{eq.rep}\\
&\bm{b}_3 = C_{3\Gamma_{1}1}\bm{\psi}_{3\Gamma_{1}1} + C_{3\Gamma_{2}1}\bm{\psi}_{3\Gamma_{2}1},+ C_{3\Gamma_{2}2}\bm{\psi}_{3\Gamma_{2}2},
\label{eq.rep_a}\\
&\bm{b}_4 = C_{4\Gamma_{1}1}\bm{\psi}_{4\Gamma_{1}1} + C_{4\Gamma_{1}2}\bm{\psi}_{4\Gamma_{1}2},+ C_{4\Gamma_{2}1}\bm{\psi}_{4\Gamma_{2}1},
\label{eq.rep_b}
\end{alignat}
where $\bm{R} = \bm{t}_n + \bm{d}$,
$\bm{t}_{n}$ and $\bm{d}$ are the $n$-th lattice position and a coordinate of the magnetic site of 
the chromium atoms (i.e., $\bm{d} = (0,0,\frac{1}{2})$, the 3b site of $R\bar{3}m$), respectively,
$C_{j\Gamma_{k}l}$ is a complex coefficient, and $\bm{\psi}_{j\Gamma_{k}l}$ is a basis vector of the irreducible 
representation for the space group $R\bar{3}m$ appearing in the $l$-th basis of a small representation 
$\Gamma_{k}$ with $\bm{q}_{j}$ (Table~\ref{table_A3}).

Least square fits were performed with six parameters of $C_{j\Gamma_{k}l}$
and show an excellent fit result with $\chi^2 = 50$.
The parameters for results at 2~K were obtained as
\begin{align}
&C_{3\Gamma_{1}1} = (0 \pm 16) {\rm{e}}^{i \phi_{3\Gamma_{1}1}},      \notag\\
&C_{3\Gamma_{2}1} = (0.15 \pm 0.01) {\rm{e}}^{i \phi_{3\Gamma_{2}1}}, \notag\\
&C_{3\Gamma_{2}2} = (0.54 \pm 0.01) {\rm{e}}^{i \phi_{3\Gamma_{2}2}}, \notag\\
&C_{4\Gamma_{1}1} =-(0.52 \pm 0.01) {\rm{e}}^{i \phi_{4\Gamma_{1}1}}, \notag\\
&C_{4\Gamma_{1}2} = (0.16 \pm 0.01) {\rm{e}}^{i \phi_{4\Gamma_{1}2}}, \notag\\
&C_{4\Gamma_{2}1} = (0.08 \pm 0.03) {\rm{e}}^{i \phi_{4\Gamma_{2}1}}. \notag
\end{align}
Note that since multiplication of a phase factor ${\rm{e}}^{i \phi_{j\Gamma_{k}l}}$
doesn't change the scattering intensity, $\phi_{j\Gamma_{k}l}$ is a certain constant.
We confirmed that obtained parameters of $C_{j\Gamma_{k}l}$
appear to be consistent with the parameters of the fitted structure of the model~(4),
leading to a non-coplanar 120$^\circ$-spin structure. 
In fact, the structure shown in Fig.~\ref{fig.5}(d) is reproduced by Eq.~(\ref{eq.rep}) with 
$C_{3\Gamma_{1}1} = -(0.06 \pm 0.01) {\rm{e}}^{i \phi_{3\Gamma_{1}1}}$,      
$C_{3\Gamma_{2}1} =  (0.15 \pm 0.01) {\rm{e}}^{i \phi_{3\Gamma_{2}1}}$, 
$C_{3\Gamma_{2}2} =  (0.53 \pm 0.01) {\rm{e}}^{i \phi_{3\Gamma_{2}2}}$, 
$C_{4\Gamma_{1}1} = -(0.52 \pm 0.01) {\rm{e}}^{i \phi_{4\Gamma_{1}1}}$,
$C_{4\Gamma_{1}2} =  (0.16 \pm 0.01) {\rm{e}}^{i \phi_{4\Gamma_{1}2}}$, and
$C_{4\Gamma_{2}1} =  (0.07 \pm 0.01) {\rm{e}}^{i \phi_{4\Gamma_{2}1}}$,
where $\phi_{3\Gamma_{1}1} = 72^{\circ} \pm 2^{\circ}$, $\phi_{3\Gamma_{2}1} = -2^{\circ} \pm2^{\circ}$, 
$\phi_{3\Gamma_{2}2} = 1^{\circ}\pm1^{\circ}$,
$\phi_{4\Gamma_{1}1} = 1^{\circ}\pm0.3^{\circ}$, 
$\phi_{4\Gamma_{1}2} = 1^{\circ}\pm4^{\circ}$, and $\phi_{4\Gamma_{2}1} = -14^{\circ}\pm2^{\circ}$, respectively.
Thus, we can now recognize that the model~(4) structure is a double-$q$ structure
consisting of all the small representations of two wave vectors.
Note that although $\phi_{j\Gamma_{k}l}$ is difficult to be determined only 
from the analysis of the scattering intensity by Eqs.~(\ref{eq.rep})--(\ref{eq.rep_b}), 
we can deduce such parameter from the fitted structure of the model (4) and its representation analysis,
as shown above.
More detailed experiments are needed to clarify precise values of $\phi_{j\Gamma_{k}l}$.
%

\begin{figure}[t]
\begin{center}
 \includegraphics[width=0.45\textwidth]{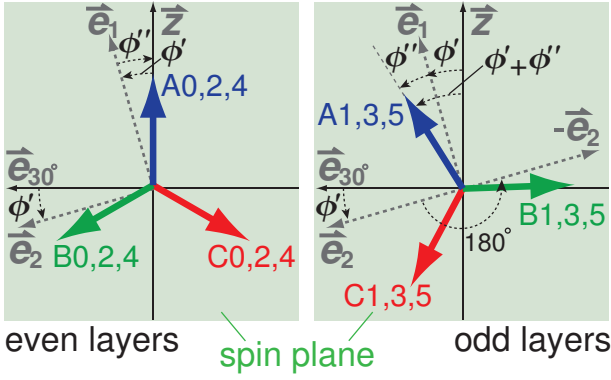}
\caption{
(Color online)
Relations of the spin configurations of the model~(3) structure
for each layer, deduced from the representation analysis. 
The rotational direction of the spins on even and odd layers is opposite,
i.e., the $+-+-+-$ structure is realized.
Definitions of $\hat{\bm e}_{1}$ and $\hat{\bm e}_{2}$ axes are in the text.
Here, $\phi' = 16^\circ$ and $\phi'' = -16^\circ$, respectively.
The $\hat{\bm c}$ axis is parallel to the $\bm{z}$ axis
and $\hat{\bm{e}}_{30^\circ}$ is the direction of the spin plane;
$\hat{\bm{e}}_{30^\circ} = \frac{\sqrt{3}}{2}\hat{{\bm x}} + \frac{1}{2}\hat{{\bm y}}$
(see text for more details).
}
\label{fig.9}
\end{center}
\end{figure}
It is worth noting here that
even though we assumed that smaller contributions of $\bm{\psi}_{3\Gamma_{1}1}$ and $\bm{\psi}_{4\Gamma_{2}1}$
were negligible and those coefficients were zero for the analysis by Eqs.~(\ref{eq.rep})--(\ref{eq.rep_b}),
we also reproduced the scattering intensities with good fittings ($\chi^2 = 51$).
This structure is a coplanar structure, corresponding to the $+-+-+-$ structure of the model~(3).
From the fittings, 
we confirmed that the magnetic structure shown in Fig.~\ref{fig.5}(c) is approximately represented 
by using parameters with 
$C_{3\Gamma_{2}1} \simeq  \frac{S}{2} \sin(\phi'){\rm e}^{i \phi''}$,
$C_{3\Gamma_{2}2} \simeq  \frac{S}{2} \cos(\phi'){\rm e}^{i \phi''}$,
$C_{4\Gamma_{1}1} \simeq -\frac{S}{2} \cos(\phi'){\rm e}^{i \phi''}$, and
$C_{4\Gamma_{1}2} \simeq  \frac{S}{2} \sin(\phi'){\rm e}^{i \phi''}$,
where $S=1.09 \pm 0.06$, $\phi' = 16^\circ \pm 1^\circ$ and $\phi'' = -16^\circ \pm 1^\circ$, respectively.
Thus, it is clear that the $+-+-+-$ structure of the model (3) is also a double-$q$ structure,
written by the linear combination of two of three small representations for each wave vector of e.g. ($\bm{q}_3, \bm{q}_4$).
In this case, Eq.~(\ref{eq.rep}) is summarized as follows:
\begin{alignat}{4}
&\bm{S}(\bm{R}) = \frac{S}{2}\cos\Bigl(\bm{q}_{3}\cdot\bm{R}-\phi''\Bigr) \hat{\bm{e}}_1 \notag\\
&\hspace{60pt}  + \frac{S}{2}\sin\Bigl(\bm{q}_{4}\cdot\bm{R}-\frac{\pi}{2}-\phi''\Bigr) \hat{\bm{e}}_2
\label{eq.rep_re}\\
&\hat{\bm{e}}_1 = \cos\Bigl(\phi'\Bigr)\hat{\bm{z}} + \sin\Bigl(\phi'\Bigr)\hat{\bm{e}}_{30^\circ} \\
&\hat{\bm{e}}_2 = \cos\Bigl(\phi'+\frac{\pi}{2}\Bigr)\hat{\bm{z}} + \sin\Bigl(\phi'+\frac{\pi}{2}\Bigr)\hat{\bm{e}}_{30^\circ}.
\end{alignat}
Calculated values of $\bm{q}_{3}\cdot\bm{R}$ and $\bm{q}_{4}\cdot\bm{R}-\pi/2$ are shown in Table~\ref{table_5}.
Since values of $\bm{q}_{4}\cdot\bm{R}-\pi/2$ are different in $\pi (= 180^\circ)$ for 
even and odd layers, the rotation direction of the spin plane becomes opposite for those layers.
This is the reason why the $+-+-+-$ structure is realized in the model (3).
Details of the spin configurations and relations between vectors of $\hat{\bm{e}}_1$ and $\hat{\bm{e}}_2$
are shown in Fig.~\ref{fig.9}.
Other results for the representation analysis of models (1) and (2)
are summarized in the appendix.
}
\begin{table}
\begin{center}
 \caption{Calculated values of $\bm{q}_{3}\cdot\bm{R}$ and $\bm{q}_{4}\cdot\bm{R}-\pi/2$ for each sublattice site.
 $i = 0,2,4$ and $j = 1,3,5$ indicate suffixes of the sublattice sites of even and odd layers, respectively.}
 \begin{tabular*}{0.48\textwidth}{@{\extracolsep{\fill}}ccc}
  \hline\hline
   \,  $\bm{R}$                  & $\bm{q}_{3}\cdot\bm{R}$   & $\bm{q}_{4}\cdot\bm{R}-\pi/2$        \rule{0mm}{3.5mm} \\ \hline
   \,  $(i = 0,2,4)$             &                           &                                      \rule{0mm}{3.5mm} \\ 
   \,  $\bm{R}_{\text{A}_i}$     &            0              &            0                         \rule{0mm}{3.5mm} \\ 
   \,  $\bm{R}_{\text{B}_i}$     &            $2\pi/3$       &            $2\pi/3$                  \rule{0mm}{3.5mm} \\ 
   \,  $\bm{R}_{\text{C}_i}$     &            $4\pi/3$       &            $4\pi/3$                  \rule{0mm}{3.5mm} \\ 
   \,  $(j = 1,3,5)$             &                           &                                      \rule{0mm}{3.5mm} \\ 
   \,  $\bm{R}_{\text{A}_j}$     &            0              &            $\pi$                     \rule{0mm}{3.5mm} \\ 
   \,  $\bm{R}_{\text{B}_j}$     &            $2\pi/3$       &            $2\pi/3 + \pi$            \rule{0mm}{3.5mm} \\ 
   \,  $\bm{R}_{\text{C}_j}$     &            $4\pi/3$       &            $4\pi/3 + \pi$            \rule{0mm}{3.5mm} \\ \hline
  \hline
 \end{tabular*}
 \label{table_5}
\end{center}
\end{table}

\section{Discussion}
\label{sec.dis}
\begin{figure}[t]
\begin{center}
 \includegraphics[width=0.45\textwidth]{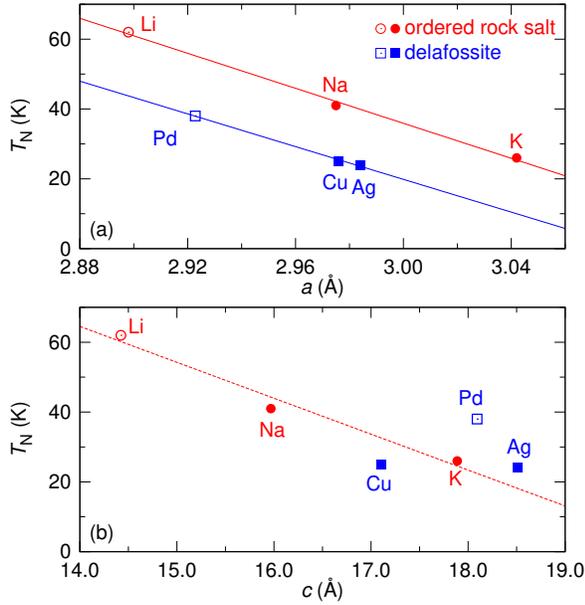}
\caption{
(Color online)
Relation between $T_{\mathrm{N}}$ and lattice parameters $a$ and $c$ of 
chromium ordered-rock salts and delafossites $A$CrO$_2$ 
(ordered rock salts: $A = \mathrm{Li}, \mathrm{Na}, \mathrm{K}$, delafossites: $A = \mathrm{Pd}, \mathrm{Cu}, \mathrm{Ag}$). 
The horizontal axis displays the lattice parameters at RT~\cite{Delmas1978-2,Soubeyroux1979,Shannon1971,Angelov1991}.
Open and filled symbols represent materials that have commensurate magnetic structures
and ones that have incommensurate magnetic structures, respectively.
In materials with smaller lattice constant $a$, 
a commensurate magnetic structure is realized probably because the nearest-neighbor interaction 
$J$ is much larger than other interactions.
}
\label{fig.8}
\end{center}
\end{figure}

For rhombohedral antiferromagnet with the delafossite type structure,
the magnetic state is expected to display a helical structure and to be highly degenerate~\cite{Rastelli1986,Rastelli1988,Reimers1992}.
The structure may become incommensurate due to certain competition among the nearest-neighbor and longer-range exchange interactions.
Experimentally, it was found recently that 
the magnetic structures of CuCrO$_2$ and AgCrO$_2$, which belong to the same magnetic Cr delafossite family as PdCrO$_2$, 
are incommensurate with a propagation vector $q = (k k 0)$ [$k\sim0.329$ for CuCrO$_2$ and $0.327$ for AgCrO$_2$].
Their magnetic structures 
are considered to be proper-screw type structures with \{110\} spiral planes~\cite{M.PoienarPRB2009,SodaJPJS2009,Oohara1994}.
In contrast, we found that
a commensurate magnetic structure is realized in PdCrO$_2$.
The transition temperature as a function of the lattice parameter $a$ reveals a linear relation in 
the delafossites $A$CrO$_2$ ($A = \mathrm{Pd}, \mathrm{Cu}, \mathrm{Ag}$) and the ordered-rock salts
$A'$CrO$_2$ ($A' = \mathrm{Li}, \mathrm{Na}, \mathrm{K}$), which all have a similar TL arrangement of Cr ions [Fig.~\ref{fig.8}(a)].
Meanwhile,
the $c$-axis length does not exhibit a simple relation with the value of $\TN$ [Fig.~\ref{fig.8}(b)].
Interestingly, 
systems that have smaller $a$ values (i.e., LiCrO$_2$ and PdCrO$_2$)
exhibit commensurate magnetic structures with magnetic Bragg reflections of ($\frac{1}{3}$ $\frac{1}{3}$ $l$) 
with $l=$ integers and half integers (i.e., a commensurate double-$q$ structure),
while systems that have larger $a$ values exhibit peculiar magnetic order including incommensurate magnetic 
orders~\cite{M.PoienarPRB2009,SodaJPJS2009,Oohara1994,Olariu2006}.
These results suggest that, while the nearest neighbor exchange interaction is predominant in all these materials,
types of other interactions are quite different between materials with smaller $a$ and those with larger $a$.

Our analysis of the magnetic structure leaves us with two possibilities for the magnetic structure of PdCrO$_2$,
the $+-+-+-$ 120$^\circ$ spin structures of the model (3) and the model (4).
The difference in the magnetic intensity between these models is too small to
distinguish them (Fig.~\ref{fig.6}).
However,
the coplanar $+-+-+-$ spin structure of the model (3)
has no scalar spin chirality and hence cannot produce an anomalous Hall current by the mechanism of the scalar spin chirality.
Even in an applied magnetic field along the $c$ axis,
the scalar spin chirality is zero because the induced spin canting would be parallel to 
the $c$ axis and hence still within the 120$^\circ$ spin plane.
In contrast,
the model (4) structure has a non-coplanar spin configuration with a locally-finite scalar spin chirality.
{\color[rgb]{0,0,0} 
Note that 
the mechanism of UAHE on the basis of the Berry-phase concept with the four-site magnetic structure~\cite{ShindouPRL2001,I.Martin2008,Y.Akagi2010} 
could be excluded for the case of PdCoO$_2$, 
because magnetic Bragg peaks are observed at $(\frac{1}{3}, \frac{1}{3},l)$ and $(\frac{2}{3}, \frac{2}{3},l)$, 
consisting of the three sublattice magnetic structure.}
We will argue now that within the model (4) structure the spin chirality mechanism could generate an UAHE.
First we note that the global sum of the scalar spin chiralities is zero in the model (4) structure.
This result implies that an additional contribution is needed to break the balance of the globally zero scalar spin chirality.
One possible mechanism is in the spin-orbit interaction~\cite{TataraJPSJ2002,Kawamura2003}.
The spin-orbit interaction breaks the perfect cancellation of local scalar spin chiralities
in the presence of a net magnetization induced by an magnetic field.
This gives then rise to an fine contribution to the UAHE~\cite{TataraJPSJ2002,Kawamura2003}.

{\color[rgb]{0,0,0} 
From the representation analysis,
the non-coplanar magnetic structure can be deduced from the symmetry properties of the $R\bar{3}m$ crystal structure of PdCrO$_2$.
In particular, the model~(4) structure is characterized by additional appearance of small representations such as 
$\Gamma_11$ of $\bm{q}_3$ and $\Gamma_21$ of $\bm{q}_4$. 
In view of the free-energy expansion, a higher-order term, 3rd or 4th, involving those small representations
becomes important for the realization of the non-coplanar double-$q$ structure,
since the simple Heisenberg type interaction with a small finite single-ion anisotropy
as given by Eq.~(\ref{eq.0}), the model (1) simplest-120$^\circ$-spin structure (i.e., single-$q$ coplanar-120$^\circ$-spin structure) 
should be realized.
More quantitative analysis is required to explain why the structure appears in PdCrO$_2$.
}

Finally, 
we discuss the small change of the magnetic structure between 2~K and 30~K.
Since the UAHE occurs in this material below $T^{*}$, we can consider two scenarios:
One is the change from the coplanar to non-coplanar spin structures 
at temperatures between 2 and 30~K.
The other scenario is the change of the amount of the non-coplanarity with temperature:
i.e.,
a non-coplanar spin structure 
is already realized at 30~K and the amount of non-coplanarity (or local scalar spin chirality) changes on cooling.
To evaluate the non-coplanarity in the magnetic structure,
we calculate the average absolute value of the local scalar spin chirality over the 18 sublattice spins,
\begin{equation}
v = \frac{\displaystyle \sum_{i,j,k}^{18} \Bigr|\bm{S}_i\cdot(\bm{S}_j \times \bm{S}_k)\Bigl|}{|M|^3},
\label{eq.10}
\end{equation}
where $M$ is the magnetic moment of each Cr spin, and $\bm{S}_i$ represents a spin component 
such as $\bm{S}_1 = \bm{S}_{\mathrm{A}0}$, $\bm{S}_2 = \bm{S}_{\mathrm{B}0}$, $\bm{S}_3 = \bm{S}_{\mathrm{C}0}$,
and etc. ($i\ne j \ne k$).
In Eq.~(\ref{eq.10}), we normalize the absolute local scalar spin chirality by $M^3$ 
in order to estimate the amount of non-coplanarity independent of the length of the ordered moment.
Although a weight of a spin-chirality contribution depends on a size of a triangle formed by three non-coplanar spins\cite{TataraJPSJ2002},
we calculate a simple sum of $\bm{S}_i\cdot(\bm{S}_j \times \bm{S}_k)$ in 18 magnetic sublattices,
without considering each the weight of each for the UAHE.
From the fit results of the model (4),
we obtained $v(2~\mathrm{K}) = 0.010 (4)$ and $v(30~\mathrm{K}) = 0.026 (4)$.
This result suggests that a slight difference appears in the non-coplanarity 
of the magnetic structures between 2~K and 30~K.
However, the value at 30~K is a bit larger than that at 2~K.
Even if the sum of chiralities given Eq.~(\ref{eq.10}) was not normalized by $M^3$, the value at 2~K is smaller.
If the second scenario is correct,
the weight of $\bm{S}_i\cdot(\bm{S}_j \times \bm{S}_k)$ from each triangle for the UAHE is different,
depending on the size of triangles or carrier mobility $\mu$:
a larger carrier mobility makes the conduction carriers interact with a larger number of spins of
the non-coplanar structure~\cite{TataraJPSJ2002}.
{\color[rgb]{0,0,0} 
This result leads to the change in the anomalous Hall conductivity $\sigma_{xy}$ originating 
in the scalar spin chirality $\chi_{ijk}$, if $\chi_{ijk}$ is finite,
because $\sigma_{xy}$ is closely related to both $\mu$ and $\chi_{ijk}$:
$\sigma_{xy} \propto \mu^2\chi_{ijk}$~[Ref.~\onlinecite{TataraJPSJ2002}].}
Experimentally, it was estimated that
the mean free path at 2~K is about ten times larger than that at 30~K~\cite{H.Takatsu2010}.
Otherwise, the first scenario may be more likely to explain the occurrence of the UAHE. 
In any case, 
a precise estimation of the knowledge of the weight of $\bm{S}_i\cdot(\bm{S}_j \times \bm{S}_k)$ 
of each triangle is important, and here the trajectory of conduction carriers should be taken into account.
The multiple-$q$ state of the magnetic structure may also be important~\cite{OkuboPRL2012}.
We now conclude that
the difference of the magnetic structure between 2~K and 30~K is small
but may result from the non-coplanarity (misfit of stacks for 120$^\circ$ layers) of the magnetic structure.
Such a minute change of the magnetic structure is consistent with a small change of 
the intensity ratio of magnetic reflections (Fig.~\ref{fig.1} inset) and of
entropy associated with a small hump in the magnetic specific heat at $T^{*}$ [Ref.~\onlinecite{H.TakatsuPRB2009}].

\section{Conclusions}
In conclusion,
we performed neutron single crystal and X-ray powder diffraction experiments
on the metallic 2D-TL antiferromagnet PdCrO$_2$ in zero field. 
We found that at 2~K the magnetic structure of Cr spins is a commensurate 120$^\circ$-spin structure 
where the spin plane includes the $c$ axis.
It alternates clockwise and anticlockwise rotation in different Cr layers.
Non-coplanar spin configurations, with an additional rotation of the spin plane 
agree with the data slightly better than the coplanar model.
In view of the observed UAHE, 
such a non-coplanar 120$^\circ$-spin structure is probably realized in PdCrO$_2$,
which is a double-$q$ structure derived by the representation analysis.
The identification of a precise non-coplanarity as well as the magnetic structure in an applied magnetic field 
should be addressed by future experiments on much larger single crystals using polarized neutrons.

\section*{Acknowledgment}
We thank T. Oguchi, S. Tatsuya, G. Tatara, R. Higashinaka, and Y. Nakai for useful discussions.
We also acknowledge the Institut Laue Langevin for beamtime allocation.
This work was supported by Grant-in-Aid for Research Activity Start-up (22840036) and 
for Young Scientists (B) (24740240) from the Ministry of Education, Culture, Sports, Science and Technology.

{\color[rgb]{0,0,0} 
\section*{Appendix}
\begin{table}
\begin{center}
 \caption{Coordinate triple of the representative symmetry operator $\{\gamma|\tau_{\gamma}\}$.}
 \begin{tabular*}{0.48\textwidth}{@{\extracolsep{\fill}}cccc}
  \hline\hline
   \,                                   & $\bm{q}_{1}$, $\bm{q}_{5}$  & $\bm{q}_{2}$, $\bm{q}_{6}$    & $\bm{q}_{3}$, $\bm{q}_{4}$      \rule{0mm}{3.5mm} \\ \hline
   \,  $\{\gamma|\tau_{\gamma}\}_1$     & $(x,y,z)$                   & $(x,y,z)$                     & $(x,y,z)$           \rule{0mm}{3.5mm} \\ 
   \,  $\{\gamma|\tau_{\gamma}\}_2$     & $(y,x,-z)$                  & $(x-y,-y,-z)$                 & $(-x,-x+y,-z)$       \rule{0mm}{3.5mm} \\ 
  \hline\hline
 \end{tabular*}
 \label{table_A1}
\end{center}
\end{table}
\begin{table}
\begin{center}
 \caption{Matrix $D_{k}(\{\gamma|\tau_{\gamma}\})$ of Eq.(\ref{eq.A1}) for small representations $\Gamma_{k}$.}
 \begin{tabular*}{0.48\textwidth}{@{\extracolsep{\fill}}ccc}
  \hline\hline
   \,  rep.             & $\{\gamma|\tau_{\gamma}\}_1$      &  $\{\gamma|\tau_{\gamma}\}_2$   \rule{0mm}{3.5mm} \\ \hline
   \,  $\Gamma_{1}$     & 1                                 &  1       \rule{0mm}{3.5mm} \\ 
   \,  $\Gamma_{2}$     & 1                                 & -1       \rule{0mm}{3.5mm} \\ 
  \hline\hline
 \end{tabular*}
 \label{table_A2}
\end{center}
\end{table}
\begin{table}
\begin{center}
 \caption{The basis vectors $\bm{\psi}_{j\Gamma_{k}l}$ of the irreducible representations of the space group $R\bar{3}m$ (point group $D^5_{3d}$) 
          appearing in the $l$-th basis of a small representation $\Gamma_{k}$ with $\bm{q}_{j}$. 
          The notation of vectors $\hat{{\bm e}}_{\alpha}$ used here is defined in the main text and the relations with other directions are 
          schematically shown in Figs.~\ref{fig.5} and \ref{fig.10}.}
 \begin{tabular*}{0.48\textwidth}{@{\extracolsep{\fill}}lcccccc}
  \hline\hline
   \,  $\bm{q}_j\, (j=1-6)$         & $\bm{q}_1$                  & $\bm{q}_2$                 & $\bm{q}_3$                  & $\bm{q}_4$                  & $\bm{q}_5$                  & $\bm{q}_6$                   \rule{0mm}{3.5mm} \\ \hline
   \,  $\bm{\psi}_{j\Gamma_{1}1}$   & $\hat{{\bm e}}_{60^\circ}$  & $\hat{{\bm e}}_{0^\circ}$  & $\hat{{\bm e}}_{-60^\circ}$ & $\hat{{\bm e}}_{30^\circ}$  & $\hat{{\bm e}}_{150^\circ}$ & $\hat{{\bm e}}_{90^\circ}$   \rule{0mm}{3.5mm} \\ 
   \,  $\bm{\psi}_{j\Gamma_{1}2}$   &                             &                            &                             & $\hat{{\bm z}}$             & $\hat{{\bm z}}$             & $\hat{{\bm z}}$              \rule{0mm}{3.5mm} \\ 
   \,  $\bm{\psi}_{j\Gamma_{2}1}$   & $\hat{{\bm e}}_{150^\circ}$ & $\hat{{\bm e}}_{90^\circ}$ & $\hat{{\bm e}}_{30^\circ}$  & $\hat{{\bm e}}_{-60^\circ}$ & $\hat{{\bm e}}_{60^\circ}$  & $\hat{{\bm e}}_{0^\circ}$    \rule{0mm}{3.5mm} \\ 
   \,  $\bm{\psi}_{j\Gamma_{2}2}$   & $\hat{{\bm z}}$             & $\hat{{\bm z}}$            & $\hat{{\bm z}}$             &                             &                             &                              \rule{0mm}{3.5mm} \\ 
  \hline\hline
 \end{tabular*}
 \label{table_A3}
\end{center}
\end{table}

\begin{table*}
\begin{center}
 \caption{Relations between the model magnetic structures and the results of the representation analysis.}
 \begin{tabular*}{0.932\textwidth}{l|l|l|l|l}
  \hline\hline
   \,\hspace{10pt}  model        \,\quad\hspace{20pt}   & \,\hspace{10pt}  type            \,\quad\hspace{30pt}  &  \,\hspace{10pt}  example                                          \,\quad\hspace{20pt}   & \,\hspace{12pt} rep.                         \,\quad\hspace{18pt}    & \,\quad\hspace{10pt}  basis vector \,\quad\hspace{21pt}                                                  \rule{0mm}{3.5mm} \\ \hline
   \,\hspace{10pt}  \multirow{2}{*}{model (1)} \,\quad  & \,\hspace{10pt}  single-$q$,     \,\quad               &  \,\hspace{10pt}  $\bm{q}_1$ domain (integer $l$)                  \,\quad                & \,\hspace{12pt} $\Gamma_{2}1$, $\Gamma_{2}2$ \,\quad                 & \,\quad\hspace{10pt}  $\bm{\psi}_{1\Gamma_{2}1}$, $\bm{\psi}_{1\Gamma_{2}2}$                             \rule{0mm}{3.5mm} \\ \cline{3-5}
   \,\hspace{10pt}                             \,\quad  & \,\hspace{10pt}  multiple domain \,\quad               &  \,\hspace{10pt}  $\bm{q}_4$ domain (half-int. $l$)                \,\quad                & \,\hspace{12pt} $\Gamma_{1}1$, $\Gamma_{1}2$ \,\quad                 & \,\quad\hspace{10pt}  $\bm{\psi}_{4\Gamma_{1}1}$, $\bm{\psi}_{4\Gamma_{1}2}$                             \rule{0mm}{3.5mm} \\ \hline
   \,\hspace{10pt}  \multirow{2}{*}{model (2)} \,\quad  & \,\hspace{10pt}  single-$q$,     \,\quad               &  \,\hspace{10pt}  $\bm{q}_1$ domain (integer $l$)                  \,\quad                & \,\hspace{12pt} $\Gamma_{2}1$, $\Gamma_{2}2$ \,\quad                 & \,\quad\hspace{10pt}  $\bm{\psi}_{1\Gamma_{2}1}$, $\bm{\psi}_{1\Gamma_{2}2}$                             \rule{0mm}{3.5mm} \\ \cline{3-5}
   \,\hspace{10pt}                             \,\quad  & \,\hspace{10pt}  multiple domain \,\quad               &  \,\hspace{10pt}  $\bm{q}_4$ domain (half-int. $l$)                \,\quad                & \,\hspace{12pt} $\Gamma_{1}1$, $\Gamma_{1}2$ \,\quad                 & \,\quad\hspace{10pt}  $\bm{\psi}_{4\Gamma_{1}1}$, $\bm{\psi}_{4\Gamma_{1}2}$                             \rule{0mm}{3.5mm} \\ \hline
   \,\hspace{10pt}  \multirow{2}{*}{model (3)} \,\quad  & \,\hspace{10pt}  double-$q$,     \,\quad               &  \,\hspace{10pt}  \multirow{2}{*}{($\bm{q}_3$,$\bm{q}_4$) domain}  \,\quad                & \,\hspace{12pt} $\Gamma_{2}1$, $\Gamma_{2}2$ \,\quad                 & \,\quad\hspace{10pt}  $\bm{\psi}_{3\Gamma_{2}1}$, $\bm{\psi}_{3\Gamma_{2}2}$                             \rule{0mm}{3.5mm} \\ 
   \,\hspace{10pt}                             \,\quad  & \,\hspace{10pt}  multiple domain \,\quad               &  \,\hspace{10pt}                                                   \,\quad                & \,\hspace{12pt} $\Gamma_{1}1$, $\Gamma_{1}2$ \,\quad                 & \,\quad\hspace{10pt}  $\bm{\psi}_{4\Gamma_{1}1}$, $\bm{\psi}_{4\Gamma_{1}2}$                             \rule{0mm}{3.5mm} \\ \hline
   \,\hspace{10pt}  \multirow{2}{*}{model (4)} \,\quad  & \,\hspace{10pt}  double-$q$,     \,\quad               &  \,\hspace{10pt}  \multirow{2}{*}{($\bm{q}_3$,$\bm{q}_4$) domain}  \,\quad                & \,\hspace{12pt} $\Gamma_{1}1$, $\Gamma_{2}1$, $\Gamma_{2}2$ \,\quad  & \,\quad\hspace{10pt}  $\bm{\psi}_{3\Gamma_{1}1}$, $\bm{\psi}_{3\Gamma_{2}1}$, $\bm{\psi}_{3\Gamma_{2}2}$ \rule{0mm}{3.5mm} \\ 
   \,\hspace{10pt}                             \,\quad  & \,\hspace{10pt}  multiple domain \,\quad               &  \,\hspace{10pt}                                                   \,\quad                & \,\hspace{12pt} $\Gamma_{1}1$, $\Gamma_{1}2$, $\Gamma_{2}1$ \,\quad  & \,\quad\hspace{10pt}  $\bm{\psi}_{4\Gamma_{1}1}$, $\bm{\psi}_{4\Gamma_{1}2}$, $\bm{\psi}_{4\Gamma_{2}1}$ \rule{0mm}{3.5mm} \\ 
  \hline\hline
 \end{tabular*}
 \label{table_A4}
\end{center}
\end{table*}
\begin{figure}
\begin{center}
 \includegraphics[width=0.45\textwidth]{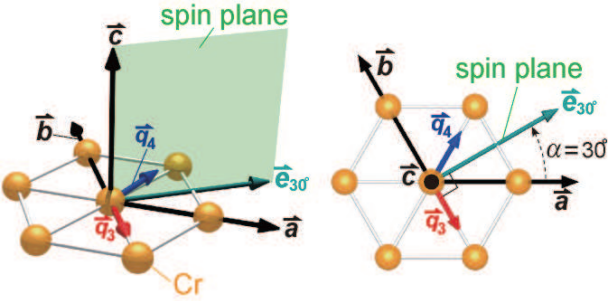}
\caption{
(Color online)
Schematic drawing of one of the relations of two wave vectors and 
the spin plane of the model (3) structure.
For $\bm{q}_3$ and $\bm{q}_4$,
the spin plane consists of $\hat{\bm{c}} ( \parallel\hat{\bm{z}})$ and $\hat{\bm{e}}_{30^\circ} (= \frac{\sqrt{3}}{2}\hat{{\bm x}} + \frac{1}{2}\hat{{\bm y}})$.
}
\label{fig.10}
\end{center}
\end{figure}

In this section, we summarize results of the representation analysis and classification of 
model magnetic structures using symmetry properties of the space group of $R\bar{3}m$, discussed in Sect.~\ref{mag.ana}.

The basis vectors $\bm{\psi}_{j\Gamma_{k}l}$ are calculated by using the projection operator method~\cite{Y.A.Izyumov}.
For the little group $G_{\bm{q}_j}$ for each wave vector $\bm{q}_{j}$, there are two one-dimensional small representations
\begin{equation}
\Gamma_{k}(\{\gamma|\tau_{\gamma}\}) = \exp(-i\bm{q}_{j}\cdot\tau_{\alpha}) D_{k}(\{\gamma|\tau_{\gamma}\}),
\label{eq.A1}
\end{equation}
where $\{\gamma|\tau_{\gamma}\}$ denotes a symmetry operator of $G_{\bm{q}_j}$
and $D_{k}(\{\gamma|\tau_{\gamma}\})$ is the matrix of the small representation of $G_{\bm{q}_j}$.
In Tables~\ref{table_A1} and \ref{table_A2}, $D_{k}(\{\gamma|\tau_{\gamma}\})$ for each $\bm{q}_{j}$ are summarized.
The basis vectors $\bm{\psi}_{j\Gamma_{k}l}$ are calculated by using these matrices and the results are listed in Table~\ref{table_A3}.

For the single-$q$ structures of models (1) and (2),
the magnetic representation of Eqs.~(\ref{eq.singleq1}), (\ref{eq.singleq2}), (\ref{eq.single_q_co_1}), 
and (\ref{eq.single_q_co_2}) can be re-written by irreducible representations: i.e., 
Eqs.~(\ref{eq.singleq1}) and (\ref{eq.single_q_co_1}) for integer-$l$ reflections are 
\begin{alignat}{2}
&\bm{S}(\bm{R}) =\bm{b}_1 \exp(i\bm{q}_{1}\cdot\bm{R}) + \bm{b}_1^* \exp(-i\bm{q}_{1}\cdot\bm{R}),\\
&\bm{b}_1 = C_{1\Gamma_{1}1}\bm{\psi}_{1\Gamma_{1}1} + C_{1\Gamma_{2}1}\bm{\psi}_{1\Gamma_{2}1},+ C_{1\Gamma_{2}2}\bm{\psi}_{1\Gamma_{2}2},
\label{eq.rep_b2}
\end{alignat}
and Eqs.~(\ref{eq.singleq2}) and (\ref{eq.single_q_co_2}) for half-integer-$l$ reflections are 
\begin{alignat}{2}
&\bm{S}(\bm{R}) =\bm{b}_4 \exp(i\bm{q}_{4}\cdot\bm{R}) + \bm{b}_4^* \exp(-i\bm{q}_{4}\cdot\bm{R}),\\
&\bm{b}_4 = C_{4\Gamma_{1}1}\bm{\psi}_{4\Gamma_{1}1} + C_{4\Gamma_{1}2}\bm{\psi}_{4\Gamma_{1}2},+ C_{4\Gamma_{2}1}\bm{\psi}_{4\Gamma_{2}1}.
\label{eq.rep_b3}
\end{alignat}
Here,
$C_{1\Gamma_{2}1} = C_{4\Gamma_{1}1} = \frac{S}{2}\exp({- \frac{\pi}{2}i})$, 
$C_{1\Gamma_{2}2} = C_{4\Gamma_{1}2} = \frac{S}{2}$, and $C_{1\Gamma_{1}l} = C_{4\Gamma_{2}l} = 0$ 
for the model~(1),
and 
$C_{1\Gamma_{2}1} = C_{4\Gamma_{1}1} = \frac{S}{2}\sin(\beta)$, 
$C_{1\Gamma_{2}2} = C_{4\Gamma_{1}2} = \frac{S}{2}\cos(\beta)$, and $C_{1\Gamma_{1}l} = C_{4\Gamma_{2}l} = 0$ 
for the model~(2), respectively.
Note that more generally,
the magnetic moment can be represented by summing the results of six-$q$ wave vectors:
\begin{alignat}{2}
&\bm{S}(\bm{R}) = \sum_{j=1}^{6} \left\{ \bm{b}_j \exp(i\bm{q}_{j}\cdot\bm{R}) + \bm{b}_j^* \exp(-i\bm{q}_{j}\cdot\bm{R})\right\},\\
&\bm{b}_X = C_{X\Gamma_{1}1}\bm{\psi}_{X\Gamma_{1}1} + C_{X\Gamma_{2}1}\bm{\psi}_{Y\Gamma_{2}1},+ C_{X\Gamma_{2}2}\bm{\psi}_{X\Gamma_{2}2},
\label{eq.rep_bX}\\
&\bm{b}_Y = C_{Y\Gamma_{1}1}\bm{\psi}_{Y\Gamma_{1}1} + C_{Y\Gamma_{1}2}\bm{\psi}_{Y\Gamma_{1}2},+ C_{Y\Gamma_{2}1}\bm{\psi}_{Y\Gamma_{2}1},
\label{eq.rep_bY}
\end{alignat}
where $X=1,2,3$ and $Y=4,5,6$, respectively.
This expression is equal to that of Eq.~(\ref{eq.magst}).
The relations of model structures and results of the representation analysis are shown in 
Table~\ref{table_A4}.
}

\bibliography{reference}

\begin{thebibliography}{54}
\expandafter\ifx\csname natexlab\endcsname\relax\def\natexlab#1{#1}\fi
\expandafter\ifx\csname bibnamefont\endcsname\relax
  \def\bibnamefont#1{#1}\fi
\expandafter\ifx\csname bibfnamefont\endcsname\relax
  \def\bibfnamefont#1{#1}\fi
\expandafter\ifx\csname citenamefont\endcsname\relax
  \def\citenamefont#1{#1}\fi
\expandafter\ifx\csname url\endcsname\relax
  \def\url#1{\texttt{#1}}\fi
\expandafter\ifx\csname urlprefix\endcsname\relax\def\urlprefix{URL }\fi
\providecommand{\bibinfo}[2]{#2}
\providecommand{\eprint}[2][]{\url{#2}}

\bibitem[{\citenamefont{Nagaosa et~al.}(2010)\citenamefont{Nagaosa, Sinova,
  Onoda, MacDonald, and Ong}}]{N.Nagaosa2010}
\bibinfo{author}{\bibfnamefont{N.}~\bibnamefont{Nagaosa}},
  \bibinfo{author}{\bibfnamefont{J.}~\bibnamefont{Sinova}},
  \bibinfo{author}{\bibfnamefont{S.}~\bibnamefont{Onoda}},
  \bibinfo{author}{\bibfnamefont{A.~H.} \bibnamefont{MacDonald}},
  \bibnamefont{and} \bibinfo{author}{\bibfnamefont{N.~P.} \bibnamefont{Ong}},
  \bibinfo{journal}{Rev. Mod. Phys.} \textbf{\bibinfo{volume}{82}},
  \bibinfo{pages}{1539} (\bibinfo{year}{2010}).

\bibitem[{\citenamefont{Xiao et~al.}(2010)\citenamefont{Xiao, Chang, and
  Niu}}]{D.Xiao2010}
\bibinfo{author}{\bibfnamefont{D.}~\bibnamefont{Xiao}},
  \bibinfo{author}{\bibfnamefont{M.-C.} \bibnamefont{Chang}}, \bibnamefont{and}
  \bibinfo{author}{\bibfnamefont{Q.}~\bibnamefont{Niu}}, \bibinfo{journal}{Rev.
  Mod. Phys.} \textbf{\bibinfo{volume}{82}}, \bibinfo{pages}{1959}
  (\bibinfo{year}{2010}).

\bibitem[{\citenamefont{Tokura and Seki}(2010)}]{TokuraAM2010}
\bibinfo{author}{\bibfnamefont{Y.}~\bibnamefont{Tokura}} \bibnamefont{and}
  \bibinfo{author}{\bibfnamefont{S.}~\bibnamefont{Seki}},
  \bibinfo{journal}{Adv. Mater.} \textbf{\bibinfo{volume}{22}},
  \bibinfo{pages}{1554} (\bibinfo{year}{2010}).

\bibitem[{\citenamefont{Arima}(2011)}]{ArimaJPSJ2011}
\bibinfo{author}{\bibfnamefont{T.}~\bibnamefont{Arima}}, \bibinfo{journal}{J.
  Phys. Soc. Jpn.} \textbf{\bibinfo{volume}{80}}, \bibinfo{pages}{052001}
  (\bibinfo{year}{2011}).

\bibitem[{\citenamefont{Taguchi et~al.}(2001)\citenamefont{Taguchi, Oohara,
  Yoshizawa, Nagaosa, and Tokura}}]{Taguchi2001}
\bibinfo{author}{\bibfnamefont{Y.}~\bibnamefont{Taguchi}},
  \bibinfo{author}{\bibfnamefont{Y.}~\bibnamefont{Oohara}},
  \bibinfo{author}{\bibfnamefont{H.}~\bibnamefont{Yoshizawa}},
  \bibinfo{author}{\bibfnamefont{N.}~\bibnamefont{Nagaosa}}, \bibnamefont{and}
  \bibinfo{author}{\bibfnamefont{Y.}~\bibnamefont{Tokura}},
  \bibinfo{journal}{Science} \textbf{\bibinfo{volume}{291}},
  \bibinfo{pages}{2573} (\bibinfo{year}{2001}).

\bibitem[{\citenamefont{Yasui et~al.}(2006)\citenamefont{Yasui, Kageyama,
  Moyoshi, Soda, Sato, and Kakurai}}]{Y.YasuiJPSJ2006}
\bibinfo{author}{\bibfnamefont{Y.}~\bibnamefont{Yasui}},
  \bibinfo{author}{\bibfnamefont{T.}~\bibnamefont{Kageyama}},
  \bibinfo{author}{\bibfnamefont{T.}~\bibnamefont{Moyoshi}},
  \bibinfo{author}{\bibfnamefont{M.}~\bibnamefont{Soda}},
  \bibinfo{author}{\bibfnamefont{M.}~\bibnamefont{Sato}}, \bibnamefont{and}
  \bibinfo{author}{\bibfnamefont{K.}~\bibnamefont{Kakurai}},
  \bibinfo{journal}{J. Phys. Soc. Jpn.} \textbf{\bibinfo{volume}{75}},
  \bibinfo{pages}{084711} (\bibinfo{year}{2006}).

\bibitem[{\citenamefont{Machida et~al.}(2007)\citenamefont{Machida, Nakatsuji,
  Maeno, Tayama, Sakakibara, and Onoda}}]{Machida2007}
\bibinfo{author}{\bibfnamefont{Y.}~\bibnamefont{Machida}},
  \bibinfo{author}{\bibfnamefont{S.}~\bibnamefont{Nakatsuji}},
  \bibinfo{author}{\bibfnamefont{Y.}~\bibnamefont{Maeno}},
  \bibinfo{author}{\bibfnamefont{T.}~\bibnamefont{Tayama}},
  \bibinfo{author}{\bibfnamefont{T.}~\bibnamefont{Sakakibara}},
  \bibnamefont{and} \bibinfo{author}{\bibfnamefont{S.}~\bibnamefont{Onoda}},
  \bibinfo{journal}{Phys. Rev. Lett.} \textbf{\bibinfo{volume}{98}},
  \bibinfo{pages}{057203} (\bibinfo{year}{2007}).

\bibitem[{\citenamefont{Machida et~al.}(2009)\citenamefont{Machida, Nakatsuji,
  Onoda, Tayama, and Sakakibara}}]{Y.MachidaNature2009}
\bibinfo{author}{\bibfnamefont{Y.}~\bibnamefont{Machida}},
  \bibinfo{author}{\bibfnamefont{S.}~\bibnamefont{Nakatsuji}},
  \bibinfo{author}{\bibfnamefont{S.}~\bibnamefont{Onoda}},
  \bibinfo{author}{\bibfnamefont{T.}~\bibnamefont{Tayama}}, \bibnamefont{and}
  \bibinfo{author}{\bibfnamefont{T.}~\bibnamefont{Sakakibara}},
  \bibinfo{journal}{Nature} \textbf{\bibinfo{volume}{463}},
  \bibinfo{pages}{210} (\bibinfo{year}{2009}).

\bibitem[{\citenamefont{Matl et~al.}(1998)\citenamefont{Matl, Ong, Yan, Li,
  Studebaker, Baum, and Doubinina}}]{P.Matl1998PRB}
\bibinfo{author}{\bibfnamefont{P.}~\bibnamefont{Matl}},
  \bibinfo{author}{\bibfnamefont{N.~P.} \bibnamefont{Ong}},
  \bibinfo{author}{\bibfnamefont{Y.~F.} \bibnamefont{Yan}},
  \bibinfo{author}{\bibfnamefont{Y.~Q.} \bibnamefont{Li}},
  \bibinfo{author}{\bibfnamefont{D.}~\bibnamefont{Studebaker}},
  \bibinfo{author}{\bibfnamefont{T.}~\bibnamefont{Baum}}, \bibnamefont{and}
  \bibinfo{author}{\bibfnamefont{G.}~\bibnamefont{Doubinina}},
  \bibinfo{journal}{Phys. Rev. B} \textbf{\bibinfo{volume}{57}},
  \bibinfo{pages}{10248} (\bibinfo{year}{1998}).

\bibitem[{\citenamefont{Ye et~al.}(1999)\citenamefont{Ye, Kim, Millis,
  Shraiman, Majumdar, and Tasanovic}}]{J.YePRL1999}
\bibinfo{author}{\bibfnamefont{J.}~\bibnamefont{Ye}},
  \bibinfo{author}{\bibfnamefont{Y.~B.} \bibnamefont{Kim}},
  \bibinfo{author}{\bibfnamefont{A.~J.} \bibnamefont{Millis}},
  \bibinfo{author}{\bibfnamefont{B.~I.} \bibnamefont{Shraiman}},
  \bibinfo{author}{\bibfnamefont{P.}~\bibnamefont{Majumdar}}, \bibnamefont{and}
  \bibinfo{author}{\bibfnamefont{Z.}~\bibnamefont{Tasanovic}},
  \bibinfo{journal}{Phys. Rev. Lett.} \textbf{\bibinfo{volume}{83}},
  \bibinfo{pages}{3737} (\bibinfo{year}{1999}).

\bibitem[{\citenamefont{Ohgushi et~al.}(2000)\citenamefont{Ohgushi, Murakami,
  and Nagaosa}}]{Ohgushi2000}
\bibinfo{author}{\bibfnamefont{K.}~\bibnamefont{Ohgushi}},
  \bibinfo{author}{\bibfnamefont{S.}~\bibnamefont{Murakami}}, \bibnamefont{and}
  \bibinfo{author}{\bibfnamefont{N.}~\bibnamefont{Nagaosa}},
  \bibinfo{journal}{Phys. Rev. B} \textbf{\bibinfo{volume}{62}},
  \bibinfo{pages}{6065} (\bibinfo{year}{2000}).

\bibitem[{\citenamefont{Tatara and Kawamura}(2002)}]{TataraJPSJ2002}
\bibinfo{author}{\bibfnamefont{G.}~\bibnamefont{Tatara}} \bibnamefont{and}
  \bibinfo{author}{\bibfnamefont{H.}~\bibnamefont{Kawamura}},
  \bibinfo{journal}{J. Phys. Soc. Jpn} \textbf{\bibinfo{volume}{71}},
  \bibinfo{pages}{2613} (\bibinfo{year}{2002}).

\bibitem[{\citenamefont{Tomizawa and Kontani}(2009)}]{T.TomizawaPRB2009}
\bibinfo{author}{\bibfnamefont{T.}~\bibnamefont{Tomizawa}} \bibnamefont{and}
  \bibinfo{author}{\bibfnamefont{H.}~\bibnamefont{Kontani}},
  \bibinfo{journal}{Phys. Rev. B} \textbf{\bibinfo{volume}{80}},
  \bibinfo{pages}{100401} (\bibinfo{year}{2009}).

\bibitem[{\citenamefont{Taguchi and Tatara}(2009)}]{K.Taguchi2009}
\bibinfo{author}{\bibfnamefont{K.}~\bibnamefont{Taguchi}} \bibnamefont{and}
  \bibinfo{author}{\bibfnamefont{G.}~\bibnamefont{Tatara}},
  \bibinfo{journal}{Phys. Rev. B} \textbf{\bibinfo{volume}{79}},
  \bibinfo{pages}{054423} (\bibinfo{year}{2009}).

\bibitem[{\citenamefont{Aharonov and Bohm}(1959)}]{Aharonov1959}
\bibinfo{author}{\bibfnamefont{Y.}~\bibnamefont{Aharonov}} \bibnamefont{and}
  \bibinfo{author}{\bibfnamefont{D.}~\bibnamefont{Bohm}},
  \bibinfo{journal}{Phys. Rev.} \textbf{\bibinfo{volume}{115}},
  \bibinfo{pages}{485} (\bibinfo{year}{1959}).

\bibitem[{\citenamefont{Takatsu et~al.}(2010)\citenamefont{Takatsu, Yonezawa,
  Fujimoto, and Maeno}}]{H.TakatsuPRL2010}
\bibinfo{author}{\bibfnamefont{H.}~\bibnamefont{Takatsu}},
  \bibinfo{author}{\bibfnamefont{S.}~\bibnamefont{Yonezawa}},
  \bibinfo{author}{\bibfnamefont{S.}~\bibnamefont{Fujimoto}}, \bibnamefont{and}
  \bibinfo{author}{\bibfnamefont{Y.}~\bibnamefont{Maeno}},
  \bibinfo{journal}{Phys. Rev. Lett.} \textbf{\bibinfo{volume}{105}},
  \bibinfo{pages}{137201} (\bibinfo{year}{2010}).

\bibitem[{\citenamefont{Shiomi et~al.}(2012)\citenamefont{Shiomi, Mochizuki,
  Kaneko, and Tokura}}]{Y.Shiomi2012}
\bibinfo{author}{\bibfnamefont{Y.}~\bibnamefont{Shiomi}},
  \bibinfo{author}{\bibfnamefont{M.}~\bibnamefont{Mochizuki}},
  \bibinfo{author}{\bibfnamefont{Y.}~\bibnamefont{Kaneko}}, \bibnamefont{and}
  \bibinfo{author}{\bibfnamefont{Y.}~\bibnamefont{Tokura}},
  \bibinfo{journal}{Phys. Rev. Lett.} \textbf{\bibinfo{volume}{108}},
  \bibinfo{pages}{056601} (\bibinfo{year}{2012}).

\bibitem[{\citenamefont{Kawamura}(2003)}]{Kawamura2003}
\bibinfo{author}{\bibfnamefont{H.}~\bibnamefont{Kawamura}},
  \bibinfo{journal}{Phys. Rev. Lett.} \textbf{\bibinfo{volume}{90}},
  \bibinfo{pages}{047202} (\bibinfo{year}{2003}).

\bibitem[{\citenamefont{Shindou and Nagaosa}(2001)}]{ShindouPRL2001}
\bibinfo{author}{\bibfnamefont{R.}~\bibnamefont{Shindou}} \bibnamefont{and}
  \bibinfo{author}{\bibfnamefont{N.}~\bibnamefont{Nagaosa}},
  \bibinfo{journal}{Phys. Rev. Lett.} \textbf{\bibinfo{volume}{87}},
  \bibinfo{pages}{116801} (\bibinfo{year}{2001}).

\bibitem[{\citenamefont{Martin and Batista}(2008)}]{I.Martin2008}
\bibinfo{author}{\bibfnamefont{I.}~\bibnamefont{Martin}} \bibnamefont{and}
  \bibinfo{author}{\bibfnamefont{C.~D.} \bibnamefont{Batista}},
  \bibinfo{journal}{Phys. Rev. Lett.} \textbf{\bibinfo{volume}{101}},
  \bibinfo{pages}{156402} (\bibinfo{year}{2008}).

\bibitem[{\citenamefont{Akagi and Motome}(2010)}]{Y.Akagi2010}
\bibinfo{author}{\bibfnamefont{Y.}~\bibnamefont{Akagi}} \bibnamefont{and}
  \bibinfo{author}{\bibfnamefont{Y.}~\bibnamefont{Motome}},
  \bibinfo{journal}{J. Phys. Soc. Jpn.} \textbf{\bibinfo{volume}{79}},
  \bibinfo{pages}{083711} (\bibinfo{year}{2010}).

\bibitem[{\citenamefont{Ok et~al.}(2013)\citenamefont{Ok, Jo, Kim, Shishidou,
  Choi, Noh, Oguchi, Min, and Kim}}]{OkPRL2013}
\bibinfo{author}{\bibfnamefont{J.~M.} \bibnamefont{Ok}},
  \bibinfo{author}{\bibfnamefont{Y.~J.} \bibnamefont{Jo}},
  \bibinfo{author}{\bibfnamefont{K.}~\bibnamefont{Kim}},
  \bibinfo{author}{\bibfnamefont{T.}~\bibnamefont{Shishidou}},
  \bibinfo{author}{\bibfnamefont{E.~S.} \bibnamefont{Choi}},
  \bibinfo{author}{\bibfnamefont{H.-J.} \bibnamefont{Noh}},
  \bibinfo{author}{\bibfnamefont{T.}~\bibnamefont{Oguchi}},
  \bibinfo{author}{\bibfnamefont{B.~I.} \bibnamefont{Min}}, \bibnamefont{and}
  \bibinfo{author}{\bibfnamefont{J.~S.} \bibnamefont{Kim}},
  \bibinfo{journal}{Phys. Rev. Lett.} \textbf{\bibinfo{volume}{111}},
  \bibinfo{pages}{176405} (\bibinfo{year}{2013}).

\bibitem[{\citenamefont{Sobota et~al.}(2013)\citenamefont{Sobota, Kim, Takatsu,
  Hashimoto, Mo, Hussain, Oguchi, Shishidou, Maeno, Min
  et~al.}}]{SobotaPRB2013}
\bibinfo{author}{\bibfnamefont{J.~A.} \bibnamefont{Sobota}},
  \bibinfo{author}{\bibfnamefont{K.}~\bibnamefont{Kim}},
  \bibinfo{author}{\bibfnamefont{H.}~\bibnamefont{Takatsu}},
  \bibinfo{author}{\bibfnamefont{M.}~\bibnamefont{Hashimoto}},
  \bibinfo{author}{\bibfnamefont{S.~K.} \bibnamefont{Mo}},
  \bibinfo{author}{\bibfnamefont{Z.}~\bibnamefont{Hussain}},
  \bibinfo{author}{\bibfnamefont{T.}~\bibnamefont{Oguchi}},
  \bibinfo{author}{\bibfnamefont{T.}~\bibnamefont{Shishidou}},
  \bibinfo{author}{\bibfnamefont{Y.}~\bibnamefont{Maeno}},
  \bibinfo{author}{\bibfnamefont{B.~I.} \bibnamefont{Min}},
  \bibnamefont{et~al.}, \bibinfo{journal}{Phys. Rev. B}
  \textbf{\bibinfo{volume}{88}}, \bibinfo{pages}{125109}
  (\bibinfo{year}{2013}).

\bibitem[{\citenamefont{Ong and Singh}(2012)}]{K.P.Ong2011}
\bibinfo{author}{\bibfnamefont{K.~P.} \bibnamefont{Ong}} \bibnamefont{and}
  \bibinfo{author}{\bibfnamefont{D.~J.} \bibnamefont{Singh}},
  \bibinfo{journal}{Phys. Rev. B} \textbf{\bibinfo{volume}{85}},
  \bibinfo{pages}{134403} (\bibinfo{year}{2012}).

\bibitem[{\citenamefont{Doumerc et~al.}(1986)\citenamefont{Doumerc,
  Wichainchai, Ammar, Pouchard, and Hagenmuller}}]{Doumerc1986}
\bibinfo{author}{\bibfnamefont{J.~P.} \bibnamefont{Doumerc}},
  \bibinfo{author}{\bibfnamefont{A.}~\bibnamefont{Wichainchai}},
  \bibinfo{author}{\bibfnamefont{A.}~\bibnamefont{Ammar}},
  \bibinfo{author}{\bibfnamefont{M.}~\bibnamefont{Pouchard}}, \bibnamefont{and}
  \bibinfo{author}{\bibfnamefont{P.}~\bibnamefont{Hagenmuller}},
  \bibinfo{journal}{Mat. Res. Bull.} \textbf{\bibinfo{volume}{21}},
  \bibinfo{pages}{745} (\bibinfo{year}{1986}).

\bibitem[{\citenamefont{Mekata et~al.}(1995)\citenamefont{Mekata, Sugino,
  Oohara, Oohara, and Yoshizawa}}]{Mekata1995}
\bibinfo{author}{\bibfnamefont{M.}~\bibnamefont{Mekata}},
  \bibinfo{author}{\bibfnamefont{T.}~\bibnamefont{Sugino}},
  \bibinfo{author}{\bibfnamefont{A.}~\bibnamefont{Oohara}},
  \bibinfo{author}{\bibfnamefont{Y.}~\bibnamefont{Oohara}}, \bibnamefont{and}
  \bibinfo{author}{\bibfnamefont{H.}~\bibnamefont{Yoshizawa}},
  \bibinfo{journal}{Physica B} \textbf{\bibinfo{volume}{213}},
  \bibinfo{pages}{221} (\bibinfo{year}{1995}).

\bibitem[{\citenamefont{Takatsu et~al.}(2009)\citenamefont{Takatsu, Yoshizawa,
  Yonezawa, and Maeno}}]{H.TakatsuPRB2009}
\bibinfo{author}{\bibfnamefont{H.}~\bibnamefont{Takatsu}},
  \bibinfo{author}{\bibfnamefont{H.}~\bibnamefont{Yoshizawa}},
  \bibinfo{author}{\bibfnamefont{S.}~\bibnamefont{Yonezawa}}, \bibnamefont{and}
  \bibinfo{author}{\bibfnamefont{Y.}~\bibnamefont{Maeno}},
  \bibinfo{journal}{Phys. Rev. B} \textbf{\bibinfo{volume}{79}},
  \bibinfo{pages}{104424} (\bibinfo{year}{2009}).

\bibitem[{\citenamefont{Hirone and Adachi}(1957)}]{T.HironeJPSJ1957}
\bibinfo{author}{\bibfnamefont{T.}~\bibnamefont{Hirone}} \bibnamefont{and}
  \bibinfo{author}{\bibfnamefont{K.}~\bibnamefont{Adachi}},
  \bibinfo{journal}{J. Phys. Soc. Jpn.} \textbf{\bibinfo{volume}{12}},
  \bibinfo{pages}{156} (\bibinfo{year}{1957}).

\bibitem[{\citenamefont{Hurd}(1972)}]{C.M.Hurd}
\bibinfo{author}{\bibfnamefont{C.~M.} \bibnamefont{Hurd}},
  \emph{\bibinfo{title}{The Hall effect in metals and alloys}}
  (\bibinfo{publisher}{Plenum Press}, \bibinfo{address}{New York},
  \bibinfo{year}{1972}).

\bibitem[{\citenamefont{Takatsu and Maeno}(2010)}]{H.Takatsu2010}
\bibinfo{author}{\bibfnamefont{H.}~\bibnamefont{Takatsu}} \bibnamefont{and}
  \bibinfo{author}{\bibfnamefont{Y.}~\bibnamefont{Maeno}}, \bibinfo{journal}{J.
  Cryst. Growth} \textbf{\bibinfo{volume}{312}}, \bibinfo{pages}{3461}
  (\bibinfo{year}{2010}).

\bibitem[{\citenamefont{Izumi and Momma}(2007)}]{Izumi2007}
\bibinfo{author}{\bibfnamefont{F.}~\bibnamefont{Izumi}} \bibnamefont{and}
  \bibinfo{author}{\bibfnamefont{K.}~\bibnamefont{Momma}},
  \bibinfo{journal}{Solid State Phenom.} \textbf{\bibinfo{volume}{130}},
  \bibinfo{pages}{15} (\bibinfo{year}{2007}).

\bibitem[{not()}]{note_cp_of_PdCrO2}
\bibinfo{note}{H. Takatsu \textit{et. al.}, unpublished}.

\bibitem[{\citenamefont{Fujiki et~al.}(1983)\citenamefont{Fujiki, Shutoh,
  yoshihiko Abe, and Katsura}}]{Fujiki1983}
\bibinfo{author}{\bibfnamefont{S.}~\bibnamefont{Fujiki}},
  \bibinfo{author}{\bibfnamefont{K.}~\bibnamefont{Shutoh}},
  \bibinfo{author}{\bibnamefont{yoshihiko Abe}}, \bibnamefont{and}
  \bibinfo{author}{\bibfnamefont{S.}~\bibnamefont{Katsura}},
  \bibinfo{journal}{J. Phys. Soc. Jpn.} \textbf{\bibinfo{volume}{52}},
  \bibinfo{pages}{1531} (\bibinfo{year}{1983}).

\bibitem[{\citenamefont{Blankschtein et~al.}(1984)\citenamefont{Blankschtein,
  Ma, and Berker}}]{Blankschtein1984}
\bibinfo{author}{\bibfnamefont{D.}~\bibnamefont{Blankschtein}},
  \bibinfo{author}{\bibfnamefont{M.}~\bibnamefont{Ma}}, \bibnamefont{and}
  \bibinfo{author}{\bibfnamefont{A.~N.} \bibnamefont{Berker}},
  \bibinfo{journal}{Phys. Rev. B} \textbf{\bibinfo{volume}{29}},
  \bibinfo{pages}{5250} (\bibinfo{year}{1984}).

\bibitem[{\citenamefont{Kimura et~al.}(2008)\citenamefont{Kimura, Nakamura,
  Ohgushi, and Kimura}}]{K.KimuraPRB2008}
\bibinfo{author}{\bibfnamefont{K.}~\bibnamefont{Kimura}},
  \bibinfo{author}{\bibfnamefont{H.}~\bibnamefont{Nakamura}},
  \bibinfo{author}{\bibfnamefont{K.}~\bibnamefont{Ohgushi}}, \bibnamefont{and}
  \bibinfo{author}{\bibfnamefont{T.}~\bibnamefont{Kimura}},
  \bibinfo{journal}{Phys. Rev. B} \textbf{\bibinfo{volume}{78}},
  \bibinfo{pages}{140401(R)} (\bibinfo{year}{2008}).

\bibitem[{sup()}]{supplement_PdCrO2}
\bibinfo{note}{Supplemental Material for the observed and calculated structure
  factors.}

\bibitem[{\citenamefont{Hemmeida et~al.}(2011)\citenamefont{Hemmeida, von
  Nidda, and Loidl}}]{Hemmeida2011}
\bibinfo{author}{\bibfnamefont{M.}~\bibnamefont{Hemmeida}},
  \bibinfo{author}{\bibfnamefont{H.-A.~K.} \bibnamefont{von Nidda}},
  \bibnamefont{and} \bibinfo{author}{\bibfnamefont{A.}~\bibnamefont{Loidl}},
  \bibinfo{journal}{J. Phys. Soc. Jpn.} \textbf{\bibinfo{volume}{80}},
  \bibinfo{pages}{053707} (\bibinfo{year}{2011}).

\bibitem[{\citenamefont{Kadowaki et~al.}(1995)\citenamefont{Kadowaki, Takei,
  and Motoya}}]{Kadowaki1995}
\bibinfo{author}{\bibfnamefont{H.}~\bibnamefont{Kadowaki}},
  \bibinfo{author}{\bibfnamefont{H.}~\bibnamefont{Takei}}, \bibnamefont{and}
  \bibinfo{author}{\bibfnamefont{K.}~\bibnamefont{Motoya}},
  \bibinfo{journal}{J. Phys. Cond. Matt.} \textbf{\bibinfo{volume}{7}},
  \bibinfo{pages}{6869} (\bibinfo{year}{1995}).

\bibitem[{\citenamefont{Brown et~al.}(2002)\citenamefont{Brown, Forsyth, and
  Tasset}}]{P.J.Brown2002}
\bibinfo{author}{\bibfnamefont{P.~J.} \bibnamefont{Brown}},
  \bibinfo{author}{\bibfnamefont{J.~B.} \bibnamefont{Forsyth}},
  \bibnamefont{and} \bibinfo{author}{\bibfnamefont{F.}~\bibnamefont{Tasset}},
  \bibinfo{journal}{J. Phys.: Cond. Matt.} \textbf{\bibinfo{volume}{14}},
  \bibinfo{pages}{1957} (\bibinfo{year}{2002}).

\bibitem[{\citenamefont{Kadowaki et~al.}(1990)\citenamefont{Kadowaki, Kikuchi,
  and Ajiro}}]{Kadowaki1990}
\bibinfo{author}{\bibfnamefont{H.}~\bibnamefont{Kadowaki}},
  \bibinfo{author}{\bibfnamefont{H.}~\bibnamefont{Kikuchi}}, \bibnamefont{and}
  \bibinfo{author}{\bibfnamefont{Y.}~\bibnamefont{Ajiro}}, \bibinfo{journal}{J.
  Phys. Cond. Matt.} \textbf{\bibinfo{volume}{2}}, \bibinfo{pages}{4485}
  (\bibinfo{year}{1990}).

\bibitem[{\citenamefont{Mekata et~al.}(1993)\citenamefont{Mekata, Yaguchi,
  Takagi, Sugino, Mituda, Yoshizawa, Hosoito, and Shinjo}}]{Mekata1993}
\bibinfo{author}{\bibfnamefont{M.}~\bibnamefont{Mekata}},
  \bibinfo{author}{\bibfnamefont{N.}~\bibnamefont{Yaguchi}},
  \bibinfo{author}{\bibfnamefont{T.}~\bibnamefont{Takagi}},
  \bibinfo{author}{\bibfnamefont{T.}~\bibnamefont{Sugino}},
  \bibinfo{author}{\bibfnamefont{S.}~\bibnamefont{Mituda}},
  \bibinfo{author}{\bibfnamefont{H.}~\bibnamefont{Yoshizawa}},
  \bibinfo{author}{\bibfnamefont{N.}~\bibnamefont{Hosoito}}, \bibnamefont{and}
  \bibinfo{author}{\bibfnamefont{T.}~\bibnamefont{Shinjo}},
  \bibinfo{journal}{J. Phys. Soc. Jpn.} \textbf{\bibinfo{volume}{62}},
  \bibinfo{pages}{4474} (\bibinfo{year}{1993}).

\bibitem[{\citenamefont{Rastelli and Tassi}(1986)}]{Rastelli1986}
\bibinfo{author}{\bibfnamefont{E.}~\bibnamefont{Rastelli}} \bibnamefont{and}
  \bibinfo{author}{\bibfnamefont{A.}~\bibnamefont{Tassi}}, \bibinfo{journal}{J.
  Phys. C: Solid State Phys.} \textbf{\bibinfo{volume}{19}},
  \bibinfo{pages}{L423} (\bibinfo{year}{1986}).

\bibitem[{\citenamefont{Rastelli and Tassi}(1988)}]{Rastelli1988}
\bibinfo{author}{\bibfnamefont{E.}~\bibnamefont{Rastelli}} \bibnamefont{and}
  \bibinfo{author}{\bibfnamefont{A.}~\bibnamefont{Tassi}}, \bibinfo{journal}{J.
  Phys. C: Solid State Phys.} \textbf{\bibinfo{volume}{21}},
  \bibinfo{pages}{1003} (\bibinfo{year}{1988}).

\bibitem[{\citenamefont{Reimers and Dahan}(1992)}]{Reimers1992}
\bibinfo{author}{\bibfnamefont{J.~N.} \bibnamefont{Reimers}} \bibnamefont{and}
  \bibinfo{author}{\bibfnamefont{J.~R.} \bibnamefont{Dahan}},
  \bibinfo{journal}{J. Phys. Cond. Matt.} \textbf{\bibinfo{volume}{4}},
  \bibinfo{pages}{8105} (\bibinfo{year}{1992}).

\bibitem[{\citenamefont{Poienar et~al.}(2009)\citenamefont{Poienar, Damay,
  Martin, Hardy, Maignan, and Andre}}]{M.PoienarPRB2009}
\bibinfo{author}{\bibfnamefont{M.}~\bibnamefont{Poienar}},
  \bibinfo{author}{\bibfnamefont{F.}~\bibnamefont{Damay}},
  \bibinfo{author}{\bibfnamefont{C.}~\bibnamefont{Martin}},
  \bibinfo{author}{\bibfnamefont{V.}~\bibnamefont{Hardy}},
  \bibinfo{author}{\bibfnamefont{A.}~\bibnamefont{Maignan}}, \bibnamefont{and}
  \bibinfo{author}{\bibfnamefont{G.}~\bibnamefont{Andre}},
  \bibinfo{journal}{Phys. Rev. B} \textbf{\bibinfo{volume}{79}},
  \bibinfo{pages}{014412} (\bibinfo{year}{2009}).

\bibitem[{\citenamefont{Soda et~al.}(2009)\citenamefont{Soda, Kimura, Kimura,
  Matsuura, and Hirota}}]{SodaJPJS2009}
\bibinfo{author}{\bibfnamefont{M.}~\bibnamefont{Soda}},
  \bibinfo{author}{\bibfnamefont{K.}~\bibnamefont{Kimura}},
  \bibinfo{author}{\bibfnamefont{T.}~\bibnamefont{Kimura}},
  \bibinfo{author}{\bibfnamefont{M.}~\bibnamefont{Matsuura}}, \bibnamefont{and}
  \bibinfo{author}{\bibfnamefont{K.}~\bibnamefont{Hirota}},
  \bibinfo{journal}{J. Phys. Soc. Jpn.} \textbf{\bibinfo{volume}{78}},
  \bibinfo{pages}{124703} (\bibinfo{year}{2009}).

\bibitem[{\citenamefont{Oohara et~al.}(1994)\citenamefont{Oohara, Mitsuda,
  Yoshizawa, Yaguchi, Kuriyama, Asano, and Mekata}}]{Oohara1994}
\bibinfo{author}{\bibfnamefont{Y.}~\bibnamefont{Oohara}},
  \bibinfo{author}{\bibfnamefont{S.}~\bibnamefont{Mitsuda}},
  \bibinfo{author}{\bibfnamefont{H.}~\bibnamefont{Yoshizawa}},
  \bibinfo{author}{\bibfnamefont{N.}~\bibnamefont{Yaguchi}},
  \bibinfo{author}{\bibfnamefont{H.}~\bibnamefont{Kuriyama}},
  \bibinfo{author}{\bibfnamefont{K.}~\bibnamefont{Asano}}, \bibnamefont{and}
  \bibinfo{author}{\bibfnamefont{M.}~\bibnamefont{Mekata}},
  \bibinfo{journal}{J. Phys. Soc. Jpn.} \textbf{\bibinfo{volume}{63}},
  \bibinfo{pages}{847} (\bibinfo{year}{1994}).

\bibitem[{\citenamefont{Olariu et~al.}(2006)\citenamefont{Olariu, Mendels,
  Bert, Ueland, Schiffer, Berger, and Cava}}]{Olariu2006}
\bibinfo{author}{\bibfnamefont{A.}~\bibnamefont{Olariu}},
  \bibinfo{author}{\bibfnamefont{P.}~\bibnamefont{Mendels}},
  \bibinfo{author}{\bibfnamefont{F.}~\bibnamefont{Bert}},
  \bibinfo{author}{\bibfnamefont{B.~G.} \bibnamefont{Ueland}},
  \bibinfo{author}{\bibfnamefont{P.}~\bibnamefont{Schiffer}},
  \bibinfo{author}{\bibfnamefont{R.~F.} \bibnamefont{Berger}},
  \bibnamefont{and} \bibinfo{author}{\bibfnamefont{R.~J.} \bibnamefont{Cava}},
  \bibinfo{journal}{Phys. Rev. Lett.} \textbf{\bibinfo{volume}{97}},
  \bibinfo{pages}{167203} (\bibinfo{year}{2006}).

\bibitem[{\citenamefont{Delmas et~al.}(1978)\citenamefont{Delmas, Menil,
  le~Flem, Fouassier, and Hagenmuller}}]{Delmas1978-2}
\bibinfo{author}{\bibfnamefont{C.}~\bibnamefont{Delmas}},
  \bibinfo{author}{\bibfnamefont{F.}~\bibnamefont{Menil}},
  \bibinfo{author}{\bibfnamefont{G.}~\bibnamefont{le~Flem}},
  \bibinfo{author}{\bibfnamefont{C.}~\bibnamefont{Fouassier}},
  \bibnamefont{and}
  \bibinfo{author}{\bibfnamefont{P.}~\bibnamefont{Hagenmuller}},
  \bibinfo{journal}{J. Phys. Chem. Solids.} \textbf{\bibinfo{volume}{39}},
  \bibinfo{pages}{51} (\bibinfo{year}{1978}).

\bibitem[{\citenamefont{Soubeyroux et~al.}(1979)\citenamefont{Soubeyroux,
  Fruchart, Dekmas, and Flem}}]{Soubeyroux1979}
\bibinfo{author}{\bibfnamefont{J.~L.} \bibnamefont{Soubeyroux}},
  \bibinfo{author}{\bibfnamefont{D.}~\bibnamefont{Fruchart}},
  \bibinfo{author}{\bibfnamefont{C.}~\bibnamefont{Dekmas}}, \bibnamefont{and}
  \bibinfo{author}{\bibfnamefont{G.~L.} \bibnamefont{Flem}},
  \bibinfo{journal}{J. Mag. Mag. Mater.} \textbf{\bibinfo{volume}{14}},
  \bibinfo{pages}{159} (\bibinfo{year}{1979}).

\bibitem[{\citenamefont{Shannon et~al.}(1971)\citenamefont{Shannon, Rogers, and
  Prewitt}}]{Shannon1971}
\bibinfo{author}{\bibfnamefont{R.~D.} \bibnamefont{Shannon}},
  \bibinfo{author}{\bibfnamefont{D.~B.} \bibnamefont{Rogers}},
  \bibnamefont{and} \bibinfo{author}{\bibfnamefont{C.~T.}
  \bibnamefont{Prewitt}}, \bibinfo{journal}{Inorg. Chem.}
  \textbf{\bibinfo{volume}{10}}, \bibinfo{pages}{713} (\bibinfo{year}{1971}).

\bibitem[{\citenamefont{Angelov and Doumerc}(1991)}]{Angelov1991}
\bibinfo{author}{\bibfnamefont{S.}~\bibnamefont{Angelov}} \bibnamefont{and}
  \bibinfo{author}{\bibfnamefont{J.~P.} \bibnamefont{Doumerc}},
  \bibinfo{journal}{Solid State Comm.} \textbf{\bibinfo{volume}{77}},
  \bibinfo{pages}{213} (\bibinfo{year}{1991}).

\bibitem[{\citenamefont{Okubo et~al.}(2012)\citenamefont{Okubo, Chung, and
  Kawamura}}]{OkuboPRL2012}
\bibinfo{author}{\bibfnamefont{T.}~\bibnamefont{Okubo}},
  \bibinfo{author}{\bibfnamefont{S.}~\bibnamefont{Chung}}, \bibnamefont{and}
  \bibinfo{author}{\bibfnamefont{H.}~\bibnamefont{Kawamura}},
  \bibinfo{journal}{Phys. Rev. Lett.} \textbf{\bibinfo{volume}{108}},
  \bibinfo{pages}{017206} (\bibinfo{year}{2012}).

\bibitem[{\citenamefont{Izyumov et~al.}(1991)\citenamefont{Izyumov, Naish, and
  Ozerov}}]{Y.A.Izyumov}
\bibinfo{author}{\bibfnamefont{Y.~A.} \bibnamefont{Izyumov}},
  \bibinfo{author}{\bibfnamefont{V.~E.} \bibnamefont{Naish}}, \bibnamefont{and}
  \bibinfo{author}{\bibfnamefont{R.~P.} \bibnamefont{Ozerov}},
  \emph{\bibinfo{title}{{Neutron Diffraction of Magnetic Materials}}}
  (\bibinfo{publisher}{Plenum Publishing}, \bibinfo{address}{New York},
  \bibinfo{year}{1991}).

\end{thebibliography}
\end{document}


\title{
Magnetic structure of the conductive triangular-lattice antiferromagnet PdCrO$_{2}$
}

\author{Hiroshi Takatsu}
\affiliation{Department of Physics, Tokyo Metropolitan University, Hachioji-shi, Tokyo 192-0397, Japan}

\author{Gwilherm N$\acute{\mathrm{e}}$nert}
\affiliation{Institut Laue-Langevin, 6 rue Jules Horowitz, BP 156,38042 Grenoble Cedex 9, France}

\author{Hiroaki Kadowaki}
\affiliation{Department of Physics, Tokyo Metropolitan University, Hachioji-shi, Tokyo 192-0397, Japan}

\author{Hideki Yoshizawa}
\affiliation{Neutron Science Laboratory, Institute for Solid State Physics, The University of Tokyo, Tokai Ibaraki 319-1106, Japan}

\author{Mechthild Enderle}
\affiliation{Institut Laue-Langevin, 6 rue Jules Horowitz, BP 156,38042 Grenoble Cedex 9, France}

\author{Shingo Yonezawa}
\affiliation{Department of Physics, Graduate School of Science, Kyoto University, Kyoto 606-8502, Japan}

\author{Yoshiteru Maeno}
\affiliation{Department of Physics, Graduate School of Science, Kyoto University, Kyoto 606-8502, Japan}

\author{Jungeun Kim}
\affiliation{Japan Synchrotron Radiation Research Institute/SPring-8, 1-1-1 Kouto, Sayo, Hyogo 679-5198, Japan}

\author{Naruki Tsuji}
\affiliation{Japan Synchrotron Radiation Research Institute/SPring-8, 1-1-1 Kouto, Sayo, Hyogo 679-5198, Japan}

\author{Masaki Takata}
\affiliation{Japan Synchrotron Radiation Research Institute/SPring-8, 1-1-1 Kouto, Sayo, Hyogo 679-5198, Japan}

\author{Yang Zhao}
\affiliation{Department of Physics and Astronomy, Johns Hopkins University, Baltimore, Maryland 21218, USA}
\affiliation{NIST Center for Neutron Research, National Institute of Standards and Technology, Gaithersburg, Maryland 20899, USA }
\affiliation{Department of Materials Science and Engineering, University of Maryland, College Park, Maryland 20742, USA}

\author{Mark Green}
\affiliation{NCNR, National Institute of Standards and Technology, Gaithersburg, MD 20899-6102, U.S.A.}

\author{Collin Broholm}
\affiliation{Department of Physics and Astronomy, Johns Hopkins University, Baltimore, Maryland 21218, USA}
\date{\today}


\pacs{75.25.-j, 72.80.Ga, 61.05.F-}
\maketitle

\section{Abstract}
In this supplemental material,
we present a list of observed and calculated nuclear and magnetic structure factors 
as well as refined parameters of magnetic structure models (1)--(4) of PdCrO$_2$.
The data were summarized for temperatures at 2 and 30~K.

\section{List of the nuclear reflections}
\begin{table*}[h]
\begin{center}
 \caption{Observed and calculated squares of the structure factor of PdCrO$_2$ collected at D10 at 2~K and 30~K. 
 The calculation was performed on the $R\bar{3}m$ structure model 
 with atomic positions Pd(0,0,0), Cr(0,0,0) and O(0,0,$z$). 
 Through the refinement, the position parameter $z$ was yielded to be $z = 0.110(1)$ at both temperatures.}
 \begin{tabular*}{0.97\textwidth}{@{\extracolsep{\fill}}lcccc}
  \hline\hline
   \,           & \multicolumn{2}{c}{2 K} & \multicolumn{2}{c}{30 K} \rule{0mm}{3.5mm} \\ \cline{2-3}\cline{4-5} 
   \, (h,k,l)   & $|F|^2_{\mathrm{obs}}$  & $|F|^2_{\mathrm{calc}}$  & $|F|^2_{\mathrm{obs}}$  & $|F|^2_{\mathrm{calc}}$   \rule{0mm}{3.5mm} \\ \hline
   \, (0,0,3)   & 1690(110)               & 1770                     &1690(110)                & 1770\,   \rule{0mm}{3.5mm} \\
   \, (0,0,6)   & 1890(130)               & 1850                     &1890(130)                & 1850\,   \rule{0mm}{3.5mm} \\
   \, (0,0,9)   & 19460(1290)             & 29680                    &19320(1290)              & 29569\,  \rule{0mm}{3.5mm} \\
   \, (0,0,12)  & 2990(200)               & 2860                     &2960(200)                & 2840\,   \rule{0mm}{3.5mm} \\
   \, (0,1,2)   & 13740(920)              & 21000                    &13630(910)               & 20900\,  \rule{0mm}{3.5mm} \\
   \, (0,1,5)   & 8710(580)               & 11730                    &8650(580)                & 11680\,  \rule{0mm}{3.5mm} \\
   \, (0,1,8)   & 24000(1600)             & 50850                    &24060(1600)              & 50650\,  \rule{0mm}{3.5mm} \\
   \, (0,1,11)  & 3540(240)               & 3650                     &3530(240)                & 3630\,   \rule{0mm}{3.5mm} \\
   \, (0,1,-1)  & 12700(850)              & 19350                    &12700(850)               & 19270\,  \rule{0mm}{3.5mm} \\
   \, (0,1,-4)  & 290(20)                 & 260                      &290(20)                  & 250\,    \rule{0mm}{3.5mm} \\
   \, (0,1,-7)  & 2330(160)               & 2370                     &2340(160)                & 2370\,   \rule{0mm}{3.5mm} \\
   \, (0,1,-10) & 26180(1750)             & 54350                    &25920(1730)              & 54100\,  \rule{0mm}{3.5mm} \\
   \, (1,1,0)   & 26980(1800)             & 69020                    &26770(1780)              & 68720\,  \rule{0mm}{3.5mm} \\
   \, (1,1,3)   & 1660(110)               & 1770                     &1650(110)                & 1770\,   \rule{0mm}{3.5mm} \\
   \, (1,1,6)   & 1860(120)               & 1850                     &1840(120)                & 1850\,   \rule{0mm}{3.5mm} \\
   \, (0,2,1)   & 10960(730)              & 19350                    &10980(730)               & 19270\,  \rule{0mm}{3.5mm} \\
   \, (0,2,-2)  & 11400(760)              & 21000                    &11260(750)               & 20900\,  \rule{0mm}{3.5mm} \\
  \hline\hline
 \end{tabular*}
 \label{table.1}
\end{center}
\end{table*}

\section{List of the magnetic reflections}
\begin{table*}[h]
\begin{center}
 \caption{Observed and calculated square of magnetic structure factors of PdCrO$_2$ at 2~K,
 collected at D10.
 Calculations were based on models (1)--(4). 
 For the model (1), results are based on a case that spins lie in the plane including the $c$ axis.
 For the model (3), 
 the calculated values are shown for the $+-+-+-$ structure, 
 which is the best fitted structure in the model (3). 
 Values of the observed structure factor were averaged values over domains due to the crystal structure symmetry.
 Here, $|F_\mathrm{M}|^2_{\mathrm{cal}}$ is represented as 
 $|F_\mathrm{M}|^2_{\mathrm{cal}}=(\frac{g}{2} f(\bm{Q}))^2 \bigr[ |F_{\mathrm{M}}(\bm{Q})|^2 \bigl]_{\mathrm{av}}$.}
 \begin{tabular*}{0.96\textwidth}{@{\extracolsep{\fill}}lccccclccccc}
  \hline\hline
                                                  & $|F_\mathrm{M}|^2_{\mathrm{obs}}$    & \multicolumn{4}{c}{$|F_\mathrm{M}|^2_{\mathrm{calc}}$}                        &                                      & $|F_\mathrm{M}|^2_{\mathrm{obs}}$    & \multicolumn{4}{c}{$|F_\mathrm{M}|^2_{\mathrm{calc}}$}      \rule{0mm}{3.5mm}\\ \cline{2-6}\cline{8-12}
   \, (h, k, l)                                   &                                      & (1)       & (2)       & (3)   & (4)    &\,\qquad (h, k, l)                                   &                                      & (1)       & (2)       & (3)    & (4)                \rule{0mm}{3.5mm} \\ \hline
   \, $(\frac{1}{3}, \frac{1}{3}, 0)$             & 365(24)                              & 223       & 352       & 342   & 343    &\,\qquad $(\frac{1}{3}, \frac{1}{3}, \frac{1}{2})$   & 134(9)                               & 174       & 130       & 131    & 129                \rule{0mm}{3.5mm} \\
   \, $(\frac{1}{3}, \frac{1}{3}, 1)$             & 359(24)                              & 150       & 347       & 336   & 341    &\,\qquad $(\frac{1}{3}, \frac{1}{3}, \frac{3}{2})$   & 322(21)                              & 254       & 316       & 323    & 320                \rule{0mm}{3.5mm} \\
   \, $(\frac{1}{3}, \frac{1}{3}, 2)$             & 203(14)                              & 139       & 202       & 193   & 196    &\,\qquad $(\frac{1}{3}, \frac{1}{3}, \frac{5}{2})$   & 88(6)                                & 150       &  82       & 87     &  84                \rule{0mm}{3.5mm} \\
   \, $(\frac{1}{3}, \frac{1}{3}, 3)$             & 213(14)                              & 163       & 208       & 203   & 205    &\,\qquad $(\frac{1}{3}, \frac{1}{3}, \frac{7}{2})$   & 230(15)                              & 132       & 224       & 225    & 224                \rule{0mm}{3.5mm} \\
   \, $(\frac{1}{3}, \frac{1}{3}, 4)$             & 208(14)                              & 107       & 204       & 202   & 205    &\,\qquad $(\frac{1}{3}, \frac{1}{3}, \frac{9}{2})$   & 243(16)                              & 149       & 233       & 238    & 236                \rule{0mm}{3.5mm} \\
   \, $(\frac{1}{3}, \frac{1}{3}, 5)$             & 49(3)                                & 91        &  51       &  49   &  50    &\,\qquad $(\frac{1}{3}, \frac{1}{3}, \frac{11}{2})$  & 97(7)                                & 96        & 100       & 104    & 103                \rule{0mm}{3.5mm} \\
   \, $(\frac{1}{3}, \frac{1}{3}, 6)$             & 68(4)                                & 88        &  75       &  75   &  76    &\,\qquad $(\frac{1}{3}, \frac{1}{3}, \frac{13}{2})$  & 168(11)                              & 79        & 170       & 172    & 171                \rule{0mm}{3.5mm} \\
   \, $(\frac{1}{3}, \frac{1}{3}, 7)$             & 80(5)                                & 61        &  82       &  83   &  84    &\,\qquad $(\frac{2}{3}, \frac{2}{3}, \frac{1}{2})$   & 42(3)                                & 80        & 46        & 47     &  46                \rule{0mm}{3.5mm} \\
   \, $(\frac{2}{3}, \frac{2}{3}, 0)$             & 156(11)                              & 103       & 162       & 157   & 158    &\,\qquad $(\frac{2}{3}, \frac{2}{3}, \frac{3}{2})$   & 156(11)                              & 123       & 147       & 150    & 149                \rule{0mm}{3.5mm} \\
   \, $(\frac{2}{3}, \frac{2}{3}, 1)$             & 142(9)                               & 70        & 141       & 136   & 138    &\,\qquad $(\frac{2}{3}, \frac{2}{3}, \frac{5}{2})$   & 75(5)                                & 72        &  71       & 72     &  71                \rule{0mm}{3.5mm} \\
   \, $(\frac{2}{3}, \frac{2}{3}, 2)$             & 149(10)                              & 66        & 153       & 148   & 150    &\,\qquad $(\frac{2}{3}, \frac{2}{3}, \frac{7}{2})$   & 31(3)                                & 65        &  31       & 33     &  32                \rule{0mm}{3.5mm} \\
   \, $(\frac{2}{3}, \frac{2}{3}, 3)$             & 134(9)                               & 85        & 127       & 124   & 124    &\,\qquad $(\frac{2}{3}, \frac{2}{3}, \frac{9}{2})$   & 116(8)                               & 86        & 112       & 115    & 114                \rule{0mm}{3.5mm} \\
   \, $(\frac{1}{3}, \frac{1}{3}, 8)$             & 9(2)                                 & 49        &  10       &  10   &  11    &\,\qquad $(\frac{1}{3}, \frac{1}{3}, \frac{17}{2})$  & 65(4)                                & 50        & 71        & 73     &  73                \rule{0mm}{3.5mm} \\
   \, $(\frac{2}{3}, \frac{2}{3}, 4)$             & 80(5)                                & 54        &  79       &  75   &  76    &\,\qquad $(\frac{2}{3}, \frac{2}{3}, \frac{11}{2})$  & 71(5)                                & 49        & 73        & 73     &  73                \rule{0mm}{3.5mm} \\
   \, $(\frac{2}{3}, \frac{2}{3}, 5)$             & 105(7)                               & 47        & 103       & 101   & 102    &\,\qquad $(\frac{1}{3}, \frac{1}{3}, \frac{19}{2})$  & 87(6)                                & 39        & 89        & 90     &  90                \rule{0mm}{3.5mm} \\
   \, $(\frac{1}{3}, \frac{1}{3}, 9)$             & 26(3)                                & 41        &  25       &  25   &  26    &\,\qquad $(\frac{2}{3}, \frac{2}{3}, \frac{13}{2})$  & 32(3)                                & 41        & 27        & 28     &  28                \rule{0mm}{3.5mm} \\
   \, $(\frac{1}{3}, \frac{1}{3},10)$             & 28(3)                                & 29        &  29       &  30   &  30    &\,\qquad $(\frac{1}{3}, \frac{4}{3},-\frac{1}{2})$   & 47(3)                                & 49        & 44        & 45     &  45                \rule{0mm}{3.5mm} \\
   \, $(\frac{2}{3}, \frac{2}{3}, 7)$             & 30(3)                                & 32        &  30       &  29   &  29    &\,\qquad $(\frac{1}{3}, \frac{4}{3}, \frac{1}{2})$   & 71(5)                                & 59        & 67        & 69     &  68                \rule{0mm}{3.5mm} \\
   \, $(\frac{1}{3}, \frac{4}{3}, 0)$             & 76(5)                                & 30        &  74       &  71   &  72    &\,\qquad $(\frac{1}{3}, \frac{4}{3},-\frac{3}{2})$   & 20(3)                                & 32        & 16        & 16     &  16                \rule{0mm}{3.5mm} \\
   \, $(\frac{1}{3}, \frac{4}{3},-1)$             & 80(5)                                & 47        &  78       &  76   &  76    &\,\qquad $(\frac{2}{3}, \frac{2}{3}, \frac{15}{2})$  & 64(4)                                & 46        & 67        & 69     &  68                \rule{0mm}{3.5mm} \\
   \, $(\frac{1}{3}, \frac{4}{3}, 2)$             & 63(4)                                & 45        &  70       &  68   &  68    &\,\qquad $(\frac{1}{3}, \frac{4}{3}, \frac{5}{2})$   & 44(3)                                & 45        & 49        & 49     &  49                \rule{0mm}{3.5mm} \\
   \, $(\frac{1}{3}, \frac{4}{3},-3)$             & 54(4)                                & 27        &  52       &  50   &  51    &\,\qquad $(\frac{1}{3}, \frac{4}{3},-\frac{5}{2})$   & 60(4)                                & 53        & 59        & 60     &  60                \rule{0mm}{3.5mm} \\
   \, $(\frac{2}{3}, \frac{2}{3}, 8)$             & 53(3)                                & 26        &  50       &  49   &  50    &\,\qquad $(\frac{1}{3}, \frac{1}{3}, \frac{21}{2})$  & 67(4)                                & 32        & 63        & 65     &  64                \rule{0mm}{3.5mm} \\
   \, $(\frac{1}{3}, \frac{4}{3},-4)$             & 65(4)                                & 37        &  60       &  58   &  59    &\,\qquad $(\frac{1}{3}, \frac{4}{3},-\frac{7}{2})$   & 38(2)                                & 40        & 34        & 35     &  34                \rule{0mm}{3.5mm} \\
   \, $(\frac{2}{3}, \frac{2}{3},10)$             & 9(3)                                 & 16        &  9        &  8    &   9    &\,\qquad $(\frac{1}{3}, \frac{4}{3}, \frac{9}{2})$   & 4(3)                                 & 25        & 3         & 4      &   3                \rule{0mm}{3.5mm} \\
   \, $(\frac{1}{3}, \frac{4}{3},-7)$             & 32(3)                                & 22        &  33       &  33   &  33    &\,\qquad $(\frac{1}{3}, \frac{4}{3},-\frac{9}{2})$   & 26(2)                                & 25        & 25        & 25     &  25                \rule{0mm}{3.5mm} \\
   \,                                             &                                      &           &           &       &        &\,\qquad $(\frac{2}{3}, \frac{2}{3}, \frac{17}{2})$  & 44(3)                                & 27        & 50        & 50     &  50                \rule{0mm}{3.5mm} \\
                                                  &                                      &           &           &       &        &\,\qquad $(\frac{1}{3}, \frac{1}{3}, \frac{23}{2})$  & 32(2)                                & 22        & 36        & 37     &  36                \rule{0mm}{3.5mm} \\
                                                  &                                      &           &           &       &        &\,\qquad $(\frac{2}{3}, \frac{2}{3}, \frac{19}{2})$  & 18(2)                                & 21        & 19        & 20     &  20                \rule{0mm}{3.5mm} \\
                                                  &                                      &           &           &       &        &\,\qquad $(\frac{1}{3}, \frac{4}{3},-\frac{13}{2})$  & 28(2)                                & 26        & 23        & 24     &  23                \rule{0mm}{3.5mm} \\
                                                  &                                      &           &           &       &        &\,\qquad $(\frac{1}{3}, \frac{1}{3}, \frac{25}{2})$  & 32(3)                                & 16        & 38        & 38     &  38                \rule{0mm}{3.5mm} \\
                                                  &                                      &           &           &       &        &\,\qquad $(\frac{1}{3}, \frac{4}{3},-\frac{15}{2})$  & 25(3)                                & 15        & 23        & 23     &  23                \rule{0mm}{3.5mm} \\
                                                  &                                      &           &           &       &        &                                                     &                                      &           &           &        &                    \rule{0mm}{-2.0mm} \\
  \hline\hline
 \end{tabular*}
 \label{table.2}
\end{center}
\end{table*}
\begin{table*}[t]
\begin{center}
 \caption{Observed and calculated square of magnetic structure factors of PdCrO$_2$ at 30~K.
 Model structures are the same as listed in Table~\ref{table.2}.}
 \begin{tabular*}{0.96\textwidth}{@{\extracolsep{\fill}}lccccclccccc}
  \hline\hline
                                                  & $|F_\mathrm{M}|^2_{\mathrm{obs}}$    & \multicolumn{4}{c}{$|F_\mathrm{M}|^2_{\mathrm{calc}}$}                                       &                                      & $|F_\mathrm{M}|^2_{\mathrm{obs}}$         & \multicolumn{4}{c}{$|F_\mathrm{M}|^2_{\mathrm{calc}}$}      \rule{0mm}{3.5mm}\\ \cline{2-6}\cline{8-12}
   \, (h, k, l)                                   &                                      & (1)       & (2)       & (3)   & (4)    &\,\qquad (h, k, l)                                   &                                      & (1)       & (2)       & (3)    & (4)               \rule{0mm}{3.5mm} \\ \hline
   \, $(\frac{1}{3}, \frac{1}{3}, 0)$             & 223(15)                              & 152       & 216       & 202   & 205    &\,\qquad $(\frac{1}{3}, \frac{1}{3}, \frac{1}{2})$   & 79(5)                                & 99        & 82        &  83    & 81                \rule{0mm}{3.5mm} \\
   \, $(\frac{1}{3}, \frac{1}{3}, 1)$             & 223(15)                              & 96        & 213       & 201   & 207    &\,\qquad $(\frac{1}{3}, \frac{1}{3}, \frac{3}{2})$   & 185(12)                              & 144       & 182       & 194    & 186               \rule{0mm}{3.5mm} \\
   \, $(\frac{1}{3}, \frac{1}{3}, 2)$             & 120(8)                               & 89        & 120       & 111   & 116    &\,\qquad $(\frac{1}{3}, \frac{1}{3}, \frac{5}{2})$   & 48(3)                                & 85        & 46        & 50     & 47                \rule{0mm}{3.5mm} \\
   \, $(\frac{1}{3}, \frac{1}{3}, 3)$             & 130(9)                               & 110       & 129       & 122   & 125    &\,\qquad $(\frac{1}{3}, \frac{1}{3}, \frac{7}{2})$   & 149(10)                              & 75        & 135       & 138    & 136               \rule{0mm}{3.5mm} \\
   \, $(\frac{1}{3}, \frac{1}{3}, 4)$             & 133(9)                               & 69        & 129       & 125   & 130    &\,\qquad $(\frac{1}{3}, \frac{1}{3}, \frac{9}{2})$   & 142(9)                               & 85        & 133       & 140    & 136               \rule{0mm}{3.5mm} \\
   \, $(\frac{1}{3}, \frac{1}{3}, 5)$             & 29(2)                                & 59        &  29       &  28   &  31    &\,\qquad $(\frac{1}{3}, \frac{1}{3}, \frac{11}{2})$  & 55(4)                                & 54        & 54        &  58    & 56                \rule{0mm}{3.5mm} \\
   \, $(\frac{1}{3}, \frac{1}{3}, 6)$             & 43(3)                                & 59        &  48       &  47   &  49    &\,\qquad $(\frac{1}{3}, \frac{1}{3}, \frac{13}{2})$  & 102(7)                               & 45        & 101       & 103    & 102               \rule{0mm}{3.5mm} \\
   \, $(\frac{1}{3}, \frac{1}{3}, 7)$             & 52(4)                                & 40        &  53       &  53   &  55    &\,\qquad $(\frac{2}{3}, \frac{2}{3}, \frac{1}{2})$   & 29(3)                                & 46        & 28        & 29     & 28                \rule{0mm}{3.5mm} \\
   \, $(\frac{2}{3}, \frac{2}{3}, 0)$             & 97(6)                                & 70        &  99       &  93   &  94    &\,\qquad $(\frac{2}{3}, \frac{2}{3}, \frac{3}{2})$   & 88(6)                                & 70        & 84        & 90     & 86                \rule{0mm}{3.5mm} \\
   \, $(\frac{2}{3}, \frac{2}{3}, 1)$             & 82(6)                                & 45        &  85       &  80   &  82    &\,\qquad $(\frac{2}{3}, \frac{2}{3}, \frac{5}{2})$   & 45(3)                                & 41        & 44        & 45     & 44                \rule{0mm}{3.5mm} \\
   \, $(\frac{2}{3}, \frac{2}{3}, 2)$             & 90(6)                                & 42        &  84       &  89   &  92    &\,\qquad $(\frac{2}{3}, \frac{2}{3}, \frac{7}{2})$   & 19(2)                                & 37        & 18        & 20     & 18                \rule{0mm}{3.5mm} \\
   \, $(\frac{2}{3}, \frac{2}{3}, 3)$             & 80(5)                                & 58        &  78       &  73   &  75    &\,\qquad $(\frac{2}{3}, \frac{2}{3}, \frac{9}{2})$   & 64(4)                                & 49        & 64        & 68     & 66                \rule{0mm}{3.5mm} \\
   \, $(\frac{1}{3}, \frac{1}{3}, 8)$             & 5(2)                                 & 32        &   6       &   7   &   8    &\,\qquad $(\frac{1}{3}, \frac{1}{3}, \frac{17}{2})$  & 34(3)                                & 29        & 39        & 41     & 40                \rule{0mm}{3.5mm} \\
   \, $(\frac{2}{3}, \frac{2}{3}, 4)$             & 44(3)                                & 35        &  47       &  43   &  45    &\,\qquad $(\frac{2}{3}, \frac{2}{3}, \frac{11}{2})$  & 43(3)                                & 28        & 45        & 45     & 45                \rule{0mm}{3.5mm} \\
   \, $(\frac{2}{3}, \frac{2}{3}, 5)$             & 79(5)                                & 30        &  64       &  61   &  64    &\,\qquad $(\frac{1}{3}, \frac{1}{3}, \frac{19}{2})$  & 49(3)                                & 22        & 52        & 54     & 53                \rule{0mm}{3.5mm} \\
   \, $(\frac{1}{3}, \frac{1}{3}, 9)$             & 19(2)                                & 27        &  16       &  16   &  18    &\,\qquad $(\frac{2}{3}, \frac{2}{3}, \frac{13}{2})$  & 18(2)                                & 23        & 15        & 16     & 15                \rule{0mm}{3.5mm} \\
   \, $(\frac{1}{3}, \frac{1}{3},10)$             & 19(3)                                & 19        &  19       &  20   &  21    &\,\qquad $(\frac{1}{3}, \frac{4}{3},-\frac{1}{2})$   & 27(3)                                & 28        & 26        & 28     & 27                \rule{0mm}{3.5mm} \\
   \, $(\frac{2}{3}, \frac{2}{3}, 7)$             & 19(2)                                & 21        &  18       &  16   &  17    &\,\qquad $(\frac{1}{3}, \frac{4}{3}, \frac{1}{2})$   & 39(3)                                & 34        & 39        & 42     & 40                \rule{0mm}{3.5mm} \\
   \, $(\frac{1}{3}, \frac{4}{3}, 0)$             & 56(4)                                & 19        &  45       &  42   &  44    &\,\qquad $(\frac{1}{3}, \frac{4}{3},-\frac{3}{2})$   & 12(3)                                & 18        & 11        & 11     & 11                \rule{0mm}{3.5mm} \\
   \, $(\frac{1}{3}, \frac{4}{3},-1)$             & 45(3)                                & 32        &  48       &  45   &  46    &\,\qquad $(\frac{2}{3}, \frac{2}{3}, \frac{15}{2})$  & 36(3)                                & 26        & 39        & 41     & 39                \rule{0mm}{3.5mm} \\
   \, $(\frac{1}{3}, \frac{4}{3}, 2)$             & 39(3)                                & 30        &  43       &  40   &  41    &\,\qquad $(\frac{1}{3}, \frac{4}{3}, \frac{5}{2})$   & 28(3)                                & 25        & 29        & 30     & 29                \rule{0mm}{3.5mm} \\
   \, $(\frac{1}{3}, \frac{4}{3},-3)$             & 34(3)                                & 17        &  31       &  29   &  30    &\,\qquad $(\frac{1}{3}, \frac{4}{3},-\frac{5}{2})$   & 37(3)                                & 30        & 34        & 36     & 35                \rule{0mm}{3.5mm} \\
   \, $(\frac{2}{3}, \frac{2}{3}, 8)$             & 29(3)                                & 17        &  31       &  30   &  32    &\,\qquad $(\frac{1}{3}, \frac{1}{3}, \frac{21}{2})$  & 34(3)                                & 18        & 36        & 38     & 37                \rule{0mm}{3.5mm} \\
   \, $(\frac{1}{3}, \frac{4}{3},-4)$             & 36(3)                                & 25        &  37       &  35   &  35    &\,\qquad $(\frac{1}{3}, \frac{4}{3},-\frac{7}{2})$   & 22(3)                                & 23        & 20        & 21     & 20                \rule{0mm}{3.5mm} \\
   \, $(\frac{2}{3}, \frac{2}{3},10)$             & 4(3)                                 & 10        &  5        &  5    &   5    &\,\qquad $(\frac{1}{3}, \frac{4}{3}, \frac{9}{2})$   & 2(2)                                 & 14        & 2         & 2      &  2                \rule{0mm}{3.5mm} \\
   \, $(\frac{1}{3}, \frac{4}{3},-7)$             & 19(3)                                & 15        &  21       &  20   &  20    &\,\qquad $(\frac{1}{3}, \frac{4}{3},-\frac{9}{2})$   & 14(3)                                & 14        & 16        & 16     & 16                \rule{0mm}{3.5mm} \\
   \,                                             &                                      &           &           &       &        &\,\qquad $(\frac{2}{3}, \frac{2}{3}, \frac{17}{2})$  & 26(3)                                & 15        & 30        & 30     & 30                \rule{0mm}{3.5mm} \\
                                                  &                                      &           &           &       &        &\,\qquad $(\frac{1}{3}, \frac{1}{3}, \frac{23}{2})$  & 17(3)                                & 12        & 20        & 21     & 20                \rule{0mm}{3.5mm} \\
                                                  &                                      &           &           &       &        &\,\qquad $(\frac{2}{3}, \frac{2}{3}, \frac{19}{2})$  & 6(2)                                 & 12        & 10        & 11     & 11                \rule{0mm}{3.5mm} \\
                                                  &                                      &           &           &       &        &\,\qquad $(\frac{1}{3}, \frac{4}{3},-\frac{13}{2})$  & 13(2)                                & 15        & 13        & 14     & 13                \rule{0mm}{3.5mm} \\
                                                  &                                      &           &           &       &        &\,\qquad $(\frac{1}{3}, \frac{1}{3}, \frac{25}{2})$  & 24(3)                                & 9         & 22        & 23     & 22                \rule{0mm}{3.5mm} \\
                                                  &                                      &           &           &       &        &\,\qquad $(\frac{1}{3}, \frac{4}{3},-\frac{15}{2})$  & 14(3)                                & 9         & 14        & 14     & 14                \rule{0mm}{3.5mm} \\
                                                  &                                      &           &           &       &        &                                                     &                                      &           &           &        &                    \rule{0mm}{-2.0mm} \\
  \hline\hline
 \end{tabular*}
 \label{table.2.2}
\end{center}
\end{table*}

\clearpage
\section{List of refined parameters}
Tables~\ref{table.3.2} and \ref{table.3.3} present results of fitting parameters. 
It is noted that we obtained two solutions in the model (2) at 2~K
and four at 30~K.
For the model (3), we obtained several solutions at both temperatures.
For the model (4), we obtained four solutions at both temperatures.
Below, we present one of the plausible results among the solutions.
We present parameters without errors,
avoiding uncertainty of those, 
since several solutions are obtained and thus exact values of errors are difficult to estimate.

\begin{table}[h]
\begin{center}
 \caption{Refined magnetic structure parameters for models (1) and (2).
Results of the fitting parameters are shown for the $q_1$ and $q_4$ wave vectors. 
Other symmetrically equivalent domains are represented by the transformation of the space group operations with
respecting 
$\bm{q}_1 \rightarrow \bm{q}_j$ ($j = 2, 3$) for integer-$l$ reflections,
and with
$\bm{q}_4 \rightarrow \bm{q}_j$ ($j = 5, 6$) for half-integer-$l$ reflections.
}
 \begin{tabular*}{0.6\textwidth}{@{\extracolsep{\fill}}crrccrr}
  \hline\hline
   \,  model(1)         & 2K        & 30K      &\,&\,  model(2)          &  2K      &\,   30K        \rule{0mm}{3.5mm} \\ \hline
   \,  $S(q_{1})$       & 0.912     & 0.740    &\,&\,  $S_{x}(q_{1})$    &  -0.18   &\,  -0.18       \rule{0mm}{3.5mm} \\ 
   \,  $S(q_{4})$       & 0.992     & 0.75     &\,&\,  $S_{y}(q_{1})$    &   0.10   &\,   0.05       \rule{0mm}{3.5mm} \\ 
   \,  $\alpha(q_{1})$  & -10       & -18      &\,&\,  $S_{z}(q_{1})$    &   0.76   &\,   0.59       \rule{0mm}{3.5mm} \\ 
   \,  $\alpha(q_{4})$  &  30       &  30      &\,&\,  $S_{x}(q_{4})$    &  -0.58   &\,  -0.52       \rule{0mm}{3.5mm} \\ 
   \,                   &           &          &\,&\,  $S_{y}(q_{4})$    &  -0.45   &\,  -0.18       \rule{0mm}{3.5mm} \\ 
   \,                   &           &          &\,&\,  $S_{z}(q_{4})$    &   0.22   &\,   0.19       \rule{0mm}{3.5mm} \\ 
  \hline\hline
 \end{tabular*}
 \label{table.3.2}
\end{center}
\end{table}

\begin{table}[h]
\begin{center}
 \caption{Refined magnetic structure parameters for models (3) and (4).
 Suffixes of the parameters for the model (4) indicate the even ($i=0,2,4$) and odd ($i=1,3,5$) layers.}
 \begin{tabular*}{0.6\textwidth}{@{\extracolsep{\fill}}crrccrr}
  \hline\hline
   \,  model(3)          & 2K     & 30K   &\,&\,  model(4)      &  2K      &\,    30K     \rule{0mm}{3.5mm} \\ \hline
   \,  $S$               & 1.10   & 0.85  &\,&\,  $S$           &  1.10    &\,   0.85     \rule{0mm}{3.5mm} \\ 
   \,  $\alpha$          & 35     & 35    &\,&\,  $\alpha_i$    &  31      &\,   27       \rule{0mm}{3.5mm} \\ 
   \,  $\phi_{1}$        & 27     & 27    &\,&\,  $\alpha_j$    &  44      &\,   51       \rule{0mm}{3.5mm} \\ 
   \,  $\phi_{2}$        &  8     & 17    &\,&\,  $\phi_{i}$    &  17      &\,   19       \rule{0mm}{3.5mm} \\ 
   \,  $\phi_{3}$        & 28     & 23    &\,&\,  $\phi_{j}$    &  16      &\,   18       \rule{0mm}{3.5mm} \\ 
   \,  $\phi_{4}$        & -9     & -5    &\,&\,  $\gamma_{i}$  &  0 (fix) &\,   0 (fix)  \rule{0mm}{3.5mm} \\ 
   \,  $\phi_{5}$        & 42     & 45    &\,&\,  $\gamma_{j}$  &  0 (fix) &\,   0 (fix)  \rule{0mm}{3.5mm} \\ 
  \hline\hline
 \end{tabular*}
 \label{table.3.3}
\end{center}
\end{table}

\begin{table}[h]
\begin{center}
 \caption{Refined parameters of complex coefficients $C_{j\Gamma_{k}l}$ of Eqs.~(18)--(20) in the main text. 
 $\phi_{j\Gamma_{k}l}$ in phase factors are fixed to be constants, since these parameters do not change the scattering intensity.}
 \begin{tabular*}{0.6\textwidth}{@{\extracolsep{\fill}}lrr}
  \hline\hline
   \,                     & 2~K ($\chi^2=50$)\hspace{16pt}                         & 30~K ($\chi^2=51$)\hspace{12pt}                                   \rule{0mm}{3.5mm} \\ \hline
   \,  $C_{3\Gamma_{1}1}$ & $(0\pm16)    {\rm{e}}^{i \phi_{3\Gamma_{1}1}}$         & $(0.03\pm0.04) {\rm{e}}^{i \phi_{3\Gamma_{1}1}}$    \rule{0mm}{3.5mm} \\ 
   \,  $C_{3\Gamma_{2}1}$ & $(0.15\pm0.01) {\rm{e}}^{i \phi_{3\Gamma_{2}1}}$       & $(0.13\pm0.01) {\rm{e}}^{i \phi_{3\Gamma_{2}1}}$    \rule{0mm}{3.5mm} \\ 
   \,  $C_{3\Gamma_{2}2}$ & $(0.54\pm0.01) {\rm{e}}^{i \phi_{3\Gamma_{2}2}}$       & $(0.42\pm0.01) {\rm{e}}^{i \phi_{3\Gamma_{2}2}}$    \rule{0mm}{3.5mm} \\ 
   \,  $C_{4\Gamma_{1}1}$ & $-(0.52\pm0.01){\rm{e}}^{i \phi_{4\Gamma_{1}1}}$       & $-(0.39\pm0.01){\rm{e}}^{i \phi_{4\Gamma_{1}1}}$    \rule{0mm}{3.5mm} \\ 
   \,  $C_{4\Gamma_{1}2}$ & $(0.16\pm0.01) {\rm{e}}^{i \phi_{4\Gamma_{1}2}}$       & $(0.14\pm0.01) {\rm{e}}^{i \phi_{4\Gamma_{1}2}}$    \rule{0mm}{3.5mm} \\ 
   \,  $C_{4\Gamma_{2}1}$ & $(0.08\pm0.03) {\rm{e}}^{i \phi_{4\Gamma_{2}1}}$       & $(0.07\pm0.02) {\rm{e}}^{i \phi_{4\Gamma_{2}1}}$    \rule{0mm}{3.5mm} \\ 
  \hline\hline
 \end{tabular*}
 \label{table.3.3}
\end{center}
\end{table}
